\documentclass{aa}  

\usepackage{graphicx}
\usepackage{mathtools, nccmath}
\usepackage[flushleft]{threeparttable} 
\usepackage{booktabs} 
\renewcommand{\arraystretch}{1.25} 
\usepackage{subcaption} 
\usepackage{multirow}
\usepackage[title]{appendix}
\usepackage{makecell}
\usepackage[varg]{txfonts}
\usepackage{url}

\usepackage{hyperref}
\hypersetup{
    colorlinks=true,
    linkcolor=blue,
    citecolor=blue,
    filecolor=blue,      
    urlcolor=blue,
    breaklinks=true,
}
\usepackage[normalem]{ulem} 

\newcommand{\vsini}{$v$\,sin\,$i_{\star}$} 
 
\newcommand{\vmic}{$v_{\rm mic}$}
\newcommand{\vmac}{$v_{\rm mac}$}

\newcommand{\teff}{$T_{\rm eff}$}
\newcommand\logg{log\,{\it g$_\star$}}
\newcommand{\met}{[M/H]}
\newcommand{\feh}{[Fe/H]}
\newcommand{\logrhk}{$\log\,\mathrm{R^\prime_{HK}}$}
\newcommand{\kms}{km\,s$^{-1}$}
\newcommand{\ms}{m~s$^{-1}$}

\newcommand{\trades}[0]{\texttt{TRADES}}
\newcommand{\pyde}[0]{\texttt{PyDE}}
\newcommand{\emcee}[0]{\texttt{emcee}}

\newcommand{\unif}[2]{\ensuremath{\mathcal{U} (#1,#2)}}

\def\aap{A\&A}

\def\apjl{ApJL}
\def\apjs{ApJS}

\defcitealias{Dransfield12022}{D22}

\bibpunct{(}{)}{;}{a}{}{,} 

\begin{document}

    \title{Mass determination of the three long-period \\Neptune- and sub-Neptune-sized planets transiting TOI-282\thanks{Based on observations performed at the European Southern Observatory under programmes 60.A-9700, 60.A-9709, 0104.C-0003, 106.21TJ.001, and 108.22EH.001.}}
    
   \author{
   A.~Barone\inst{\ref{inst:1}, \ref{inst:3}}\,$^{\href{https://orcid.org/0009-0001-5140-8220}{\protect\includegraphics[height=0.22cm]{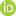}}}$
   \and 
   F.~Rodler\inst{\ref{inst:2}}\,$^{\href{https://orcid.org/0000-0003-0650-5723}{\protect\includegraphics[height=0.22cm]{figures/orcid.png}}}$
   \and
   D.~Gandolfi\inst{\ref{inst:1}}\,$^{\href{https://orcid.org/0000-0001-8627-9628}{\protect\includegraphics[height=0.22cm]{figures/orcid.png}}}$
   \and 
   A.~Bonfanti\inst{\ref{inst:3}}\,$^{\href{https://orcid.org/0000-0002-1916-5935}{\protect\includegraphics[height=0.22cm]{figures/orcid.png}}}$
   \and 
   P.~Leonardi\inst{\ref{inst:4},\ref{inst:5}}\,$^{\href{https://orcid.org/0000-0001-6026-9202}{\protect\includegraphics[height=0.22cm]{figures/orcid.png}}}$
   \and 
   L.~Visca\inst{\ref{inst:1}}
   \and
   M.~Fridlund\inst{\ref{inst:6},\ref{inst:7}}\,$^{\href{https://orcid.org/0000-0002-0855-8426}{\protect\includegraphics[height=0.22cm]{figures/orcid.png}}}$ 
   \and
   M.~Brogi\inst{\ref{inst:1},\ref{inst:8}}\,$^{\href{https://orcid.org/0000-0002-7704-0153}{\protect\includegraphics[height=0.22cm]{figures/orcid.png}}}$
   \and
   L.~Fossati\inst{\ref{inst:3}}\,$^{\href{https://orcid.org/0000-0003-4426-9530}{\protect\includegraphics[height=0.22cm]{figures/orcid.png}}}$
   \and 
   P.~E.~Cubillos\inst{\ref{inst:3}, \ref{inst:8}}\,$^{\href{https://orcid.org/0000-0002-1347-2600}{\protect\includegraphics[height=0.22cm]{figures/orcid.png}}}$
   \and W.~D.~Cochran\inst{\ref{inst:9},\ref{inst:10}}\,$^{\href{https://orcid.org/0000-0001-9662-3496}{\protect\includegraphics[height=0.22cm]{figures/orcid.png}}}$ 
   \and 
    S.~Csizmadia\inst{\ref{inst:16}}\,$^{\href{https://orcid.org/0000-0001-6803-9698}
   {\protect\includegraphics[height=0.22cm]{figures/orcid.png}}}$
   \and
   J.~Livingston\inst{\ref{inst:17},\ref{inst:18},\ref{inst:19}}\,$^{\href{https://orcid.org/0000-0002-4881-3620}{\protect\includegraphics[height=0.22cm]{figures/orcid.png}}}$
   \and
   G.~Nowak\inst{\ref{inst:11}}\,$^{\href{https://orcid.org/0000-0002-7031-7754}{\protect\includegraphics[height=0.22cm]{figures/orcid.png}}}$
   \and 
   E.~Pallé\inst{\ref{inst:12},\ref{inst:13}}\,$^{\href{https://orcid.org/0000-0003-0987-1593}{\protect\includegraphics[height=0.22cm]{figures/orcid.png}}}$
   \and 
   C.~M.~Persson\inst{\ref{inst:7}}\,$^{\href{https://orcid.org/0000-0003-1257-5146}{\protect\includegraphics[height=0.22cm]{figures/orcid.png}}}$
   \and
   S.~Redfield\inst{\ref{inst:14}}\,$^{\href{https://orcid.org/0000-0003-3786-3486}{\protect\includegraphics[height=0.22cm]{figures/orcid.png}}}$
   \and    
   H.~Schmerling\inst{\ref{inst:15}}
   \and
   A.~M.~S.~Smith\inst{\ref{inst:16}}\,$^{\href{https://orcid.org/0000-0002-2386-4341}{\protect\includegraphics[height=0.22cm]{figures/orcid.png}}}$ 
   }

   \institute{
   \label{inst:1} Dipartimento di Fisica, Università degli Studi di Torino, via Pietro Giuria 1, I-10125, Torino, Italy 
   \and
   \label{inst:2} European Southern Observatory, Alonso de Cordova 3107, Vitacura, Santiago de Chile, Chile 
   \and
   \label{inst:3} Space Research Institute, Austrian Academy of Sciences, Schmiedlstra{\ss}e 6, A-8042 Graz, Austria 
   \and
   \label{inst:4} Dipartimento di Fisica, Università di Trento, Via Sommarive 14, 38123 Povo 
   \and
   \label{inst:5} Dipartimento di Fisica e Astronomia ``Galileo Galilei'', Università degli Studi di Padova, Vicolo dell'Osservatorio 3, 35122 Padova, Italy 
   \and
   \label{inst:6} Leiden Observatory, University of Leiden, PO Box 9513, 2300 RA Leiden, The Netherlands 
   \and
   \label{inst:7} Department of Space, Earth and Environment, Chalmers University of Technology, Onsala Space Observatory, 439 92 Onsala, Sweden 
   \and
   \label{inst:8} Osservatorio Astrofisico di Torino, INAF, via Osservatorio 20, Pino Torinese, I-10025, Italy
   \and
   \label{inst:9} McDonald Observatory and Department of Astronomy, The University of Texas at Austin, USA 
   \and
   \label{inst:10} Center for Planetary Systems Habitability, The University of Texas at Austin, USA 
   \and
   \label{inst:11} Institute of Astronomy, Faculty of Physics, Astronomy and Informatics, Nicolaus Copernicus University, Grudzi\c{a}dzka 5, 87-100 Toru\'n, Poland 
   \and
   \label{inst:12} Instituto de Astrofísica de Canarias, Vía Láctea s/n, 38200 La Laguna, Tenerife, Spain 
   \and
   \label{inst:13} Departamento de Astrofísica, Universidad de La Laguna, Astrofísico Francisco Sanchez s/n, 38206 La Laguna, Tenerife, Spain \and
   \label{inst:14} Astronomy Department and Van Vleck Observatory, Wesleyan University, Middletown, CT 06459, USA 
   \and
   \label{inst:15} Rheinisches Institut f\"ur Umweltforschung an der Universit\"at zu K\"oln, Aachener Stra{\ss}e 209, 50931 K\"oln, Germany 
   \and
   \label{inst:16} Institute of Space Research, German Aerospace Center (DLR), Rutherfordstra{\ss}e 2, 12489 Berlin, Germany 
   \and
   \label{inst:17} Astrobiology Center, NINS, 2-21-1 Osawa, Mitaka, Tokyo 181-8588, Japan
   \and
   \label{inst:18} National Astronomical Observatory of Japan, NINS, 2-21-1 Osawa, Mitaka, Tokyo 181-8588, Japan
   \and
   \label{inst:19} Astronomical Science Program Graduate University for Advanced Studies, SOKENDAI, 2-21-1, Osawa, Mitaka, Tokyo, 181-8588, Japan
   }

   \date{Received 08 Aug 2025 / Accepted 29 Oct 2025}
 
  \abstract
  {TOI-282 is a bright (V\,=\,9.38) F8 main-sequence star known to host three transiting long-period ($P_b$\,=\,22.9~d, $P_c$\,=\,56.0~d, and $P_d$\,=\,84.3~d) small ($R_p$\,$\approx$\,2\,-\,4~$R_{\oplus}$) planets. The orbital period ratio of the two outermost planets, namely TOI-282\,c and d, is close to the 3:2 commensurability, suggesting that the planets might be trapped in a mean motion resonance. We combined space-borne photometry from the TESS telescope with high-precision HARPS and ESPRESSO Doppler measurements to refine orbital parameters, measure the planetary masses, and investigate the architecture and evolution of the system. We performed a Markov chain Monte Carlo joint analysis of the transit light curves and radial velocity time series, and carried out a dynamical analysis to model transit timing variations and Doppler measurements along with N-body integration. In agreement with previous results, we found that TOI-282\,b, c, and d have radii of $R_b=2.69 \pm 0.23 \ R_{\oplus}$, $R_c=4.13^{+0.16}_{-0.14} \ R_{\oplus}$, and $R_d=3.11 \pm 0.15 \ R_{\oplus}$, respectively. We measured planetary masses of $M_b=6.2\pm1.6 \ M_{\oplus}$, $M_c=9.2\pm2.0 \ M_{\oplus}$, and $M_d=5.8^{+0.9}_{-1.1} \ M_{\oplus}$, which imply mean densities of $\rho_b=1.8^{+0.7}_{-0.6} \ \text{g cm}^{-3}$, $\rho_c=0.7 \pm 0.2 \ \text{g cm}^{-3}$, and $\rho_d=1.1^{+0.3}_{-0.2} \ \text{g cm}^{-3}$, respectively. The three planets may be water worlds, making TOI-282 an interesting system for future atmospheric follow-up observations with JWST and ELT.}


\maketitle

\nolinenumbers

\section{Introduction} \label{sec: Introduction}

Multi-planet systems are remarkable laboratories for investigating the nature of planets orbiting stars other than the Sun. They provide a unique opportunity to 'freeze' the age and chemical composition of the star and its planet-forming disk, which are crucial parameters to study planetary formation and evolution. This enables direct comparisons among the planets within a single system, enabling the investigation of their architecture and mutual gravitational interactions, ultimately providing valuable insights into their formation and evolutionary pathways \citep[e.g.][]{Petrovich2020, Weiss2022, Mishra2023, Leleu2024, Howe2025}. We can also compare different planetary systems to study patterns as a function of their parameters, such as stellar spectral type, age, and instellation \citep[e.g.][]{Gupta2020, Spaargaren2020, Luque2022, Rogers2023, Venturini2024}. This is only possible by precisely and accurately characterising the planets within the system. Deriving radii and masses of planets, and consequently bulk densities, allows us to study their internal structure and composition, providing important hints about their formation and evolution \citep{Winn2015}.

Among the various types of exoplanets discovered to date, super-Earths (1\,$\lesssim$\,$R_\mathrm{p}$\,$\lesssim$ 2~$R_{\oplus}$) and sub-Neptunes (2\,$\lesssim$\, $R_\mathrm{p}$\,$\lesssim$\,4~$R_{\oplus}$) stand out as the most intriguing ones. These planets are not present in our Solar System, yet they are found orbiting around nearly half of the solar-like stars in the Galaxy \citep{Batalha2013, Petigura2013, Marcy2014, Zang2025}. Investigating their nature and properties is of particular interest, as it may help us to put into context the apparent uniqueness of our Solar System and further explore their peculiar demographics.

Long-period ($P$\,$\gtrsim$\,50~d) transiting sub-Neptunes offer a unique opportunity to investigate the properties of exoplanet atmospheres that remain largely unaffected by tidal interactions with their host stars and/or by stellar irradiation. Residing in cooler environments, their chemical and dynamical processes may resemble those of the Solar System's planets, making them compelling targets for exploring atmospheric diversity \citep{Giacobbe2021}.  Moreover, transiting long-period sub-Neptunes are ideal candidates for atmospheric characterisation with the James Webb Space Telescope \citep[JWST; ][]{Gardner2006}, which might even lead to the discovery of biosignatures \citep[e.g.][]{Madhusudhan2016, Madhusudhan2019, Kempton2024}. 

Although the number of confirmed exoplanets has increased to more than 6000\footnote{As of September 2025; source: \href{https://exoplanetarchive.ipac.caltech.edu}{NASA Exoplanet Archive}.}, only $\sim$20\% of these have orbital periods exceeding 50~d. Narrowing down to long-period transiting sub-Neptunes ($P$\,$\gtrsim$\,50~d and 2\,$\lesssim$\, $R_\mathrm{p}$\,$\lesssim$\,4~$R_{\oplus}$), this percentage falls to $\sim$5\%, due to the well-known observational biases of the radial velocity and transit methods \citep{Winn2015}. Characterising the existing relatively small sample of transiting long-period sub-Neptunes represents a key step towards comprehending their formation and evolution processes.

Here we present the characterisation of the multi-planet system \object{TOI-282} (aka \object{HD\,28109}), a bright (V\,=\,9.38), main-sequence star of spectral type F8 known to host three transiting long-period Neptunian planets, namely \object{TOI-282\,b} ($P_\mathrm{b}$\,$\approx$\,22.9\,d, $R_\mathrm{b}$\,$\approx$\,2.2\,$R_\oplus$), \object{TOI-282\,c} ($P_\mathrm{c}$\,$\approx$\,56.0\,d, $R_\mathrm{c}$\,$\approx$\,4.2\,$R_\oplus$), and \object{TOI-282\,d} ($P_\mathrm{d}$\,$\approx$\,84.3\,d, $R_\mathrm{d}$\,$\approx$\,3.3\,$R_\oplus$). The discovery of the system has been announced by \citealt{Dransfield12022} (hereafter \citetalias{Dransfield12022}), who combined TESS space-borne photometry with ground-based transit observations to determine the orbital period and radius of each planet. With the aim of measuring the masses of the three planets, we carried out intensive radial velocity (RV) follow-up observations of the star using the HARPS and ESPRESSO spectrographs and jointly modelled the Doppler measurements with TESS transit photometry. 

The paper is organised as follows. Sect.~\ref{sec: Observations and data reduction} presents the photometric and spectroscopic data. Sect.~\ref{sec: Stellar characterization} describes our determination of the fundamental parameters of the star TOI-282. Sect.~\ref{sec: Frequency analysis} outlines the frequency analysis of the RV and activity indicator time series. After describing the data modelling in Sect.~\ref{sec: Methods}, we present the planetary parameters (Sect.~\ref{sec: LC and RV data analysis results}) and a dynamical analysis (Sect.~\ref{sec: Dynamical analysis}). Finally, discussions and conclusions are provided in Sects.~\ref{sec: The planetary system TOI-282} and \ref{sec: Conclusions}, respectively.

\section{Observations and data reduction} \label{sec: Observations and data reduction}

\subsection{TESS photometry} \label{ssec: TESS photometry}

The photometric analysis presented in this paper is exclusively based on publicly available TESS \citep{Ricker2015} space-borne photometry, which we retrieved from the STScI Mikulski Archive for Space Telescope\footnote{\url{https://mast.stsci.edu/portal/Mashup/Clients/Mast/Portal.html}.} (MAST). As of June 2025, TESS has observed TOI-282 in 37 sectors, namely, sectors 1-13, 27-31, 33-39, 61-69, and 87-90. The TESS observations used in the present paper cover a baseline of nearly 7 years, from July 2018 to March 2025. The results presented in \citetalias{Dransfield12022} are based on the first 23 TESS sectors (July 2018-April 2021; sectors 1-13, 27-31, and 33-37).

We downloaded the two-minute cadence light curves (LCs) as processed and calibrated by the Science Processing Operations Center \citep[SPOC;][]{Jenkins2016} at NASA Ames Research Center. For our analysis, we used the data products delivered in the form of Presearch Data Conditioning Simple Aperture Photometry \citep[PDCSAP;][]{Smith2012, Stumpe2012}, which is corrected for instrumental artefacts, systematic trends, and contaminations. We discarded those data points with poor quality flags and extracted $\sim$4 hours of out-of-transit photometry both before and after each transit for de-trending purposes. Our photometric data set comprises 59 transit LCs, which are broken down as follows: 33, 18, and 8 transits for TOI-282\,b, c, and d, respectively. The 59 transit LCs contain time stamps in barycentric Julian dates in barycentric dynamical time \citep[BJD$_\mathrm{TDB}$;][]{Eastman2010}, along with the corresponding PDCSAP fluxes and uncertainties. The time series includes the following ancillary vectors that are part of the TESS science data product: $\texttt{mom\_centr1}$, $\texttt{mom\_centr2}$, $\texttt{pos\_corr1}$, and $\texttt{pos\_corr2}$.

\subsection{HARPS spectroscopy} 
\label{sec: HARPS high-resolution spectroscopy}

We started the RV follow-up of TOI-282 using the High Accuracy Radial Velocity Planet Searcher \citep[HARPS;][]{Mayor2003} echelle spectrograph installed on the ESO 3.6-m telescope at La Silla Observatory, Chile. Two high-resolution (R\,$\approx$\,115\,000) spectra of TOI-282 were acquired on 21 and 22 February 2019 (program ID 60.A-9700). We collected 36 additional spectra between 6 December 2020 and 17 July 2021 (program IDs 60.A-9709 and 106.21TJ.001; PI: D. Gandolfi). The observations were carried out as part of our intensive RV follow-up program of TESS small transiting planets conducted with the HARPS spectrograph \citep[see, e.g.][]{Gandolfi2025}. The exposure time was set to 1500\,-\,1800~s, which resulted in a median signal-to-noise ratio (S/N) of $\sim$104 per pixel at 550 nm. One spectrum was discarded due to contamination from scattered moonlight, leading to 37 useful HARPS spectra. 

We reduced the spectra using the dedicated HARPS data reduction software \citep[\texttt{DRS};][]{Lovis2007} and computed the cross-correlation functions (CCFs) using a G2 numerical mask \citep{Baranne1996, Pepe2002}. We also utilised the \texttt{DRS} to extract the \logrhk\ chromospheric index from the Ca\,{\sc ii}\,H\,\&\,K lines (Table~\ref{table:HARPS_ActivityIndicators_Table}). Following the methodology described in \cite{Simola2019}, we fitted a skew-normal (SN) function to the HARPS CCFs to extract the RV measurements along with three profile diagnostics of the SN function, namely, the full width at half maximum FWHM, the contrast $A$, and the skewness $\gamma$ (Table~\ref{table:HARPS_RV_Table}). The RV uncertainties $\sigma_{\text{RV}}$ were derived from the variance-covariance matrix of the model parameters following \cite{Richter1995}. We finally extracted the Na\,D1, Na\,D2, and H$_{\alpha}$ lines activity indicators (Table~\ref{table:HARPS_ActivityIndicators_Table}) using the Template Enhanced Radial Velocity Re-analysis Application \citep[\texttt{TERRA};][]{Anglada2012}.

\subsection{ESPRESSO spectroscopy} 
\label{ssec: ESPRESSO spectroscopic data}
 
We also observed TOI-282 using the Echelle SPectrograph for Rocky Exoplanet and Stable Spectroscopic Observations \citep[ESPRESSO;][]{Pepe2014, Pepe2021} installed on the 8.2-m ESO Very Large Telescope (VLT) at Paranal Observatory, Chile. The observations were carried out as part of two dedicated ESPRESSO programs to spectroscopically follow up TOI-282 (program IDs 0104.C-0003 and 108.22EH.001; PI: F.~Rodler). We acquired 44 high-resolution (R\,$\approx$\,140\,000) spectra covering a baseline of nearly 2.5 years, between October 2019 and March 2022. We set the exposure time to 935\,-\,1270 s, achieving a median S/N of 122 per pixel at 550~nm. In November and December 2020, ESPRESSO underwent an upgrade of the Fabry-Pérot lamp. To account for a possible RV offset resulting from the instrument refurbishment, we split the data into ESPRESSO 1st and 2nd sets. The former is composed of 8 spectra acquired between 15 October and 27 December 2019, whereas the latter includes the remaining 36 spectra that were secured between 5 October 2021 and 2 March 2022. As performed with HARPS data (Sect.~\ref{sec: HARPS high-resolution spectroscopy}), we reduced the ESPRESSO spectra with the dedicated \texttt{DRS} \citep{Pepe2021}, computed the CCFs using a F9 mask, and extracted the RV measurements and profile diagnostics by fitting an SN function to the ESPRESSO CCFs (Table~\ref{table:ESPRESSO_RV_Table}).

\section{Stellar characterisation}
\label{sec: Stellar characterization}

\begin{table}
\caption{Main identifiers, equatorial coordinates, parallax, distance, optical and near-infrared magnitudes, and fundamental parameters of the star TOI-282.}
\label{tab:star}
\centering
\begin{tabular}{l r c}
\hline\hline   
\noalign{\smallskip}
\multicolumn{1}{l}{\multirow{5}*{Main identifiers}} & \multicolumn{2}{l}{TOI-282} \\
\multicolumn{1}{l}{}               & \multicolumn{2}{l}{TIC 178155732} \\
\multicolumn{1}{l}{}               & \multicolumn{2}{l}{HD 28109} \\
\multicolumn{1}{l}{}               & \multicolumn{2}{l}{HIP 20295} \\
\multicolumn{1}{l}{}               & \multicolumn{2}{l}{Gaia DR3 4668163021600295552} \\
\noalign{\smallskip}
\hline
\noalign{\smallskip}
Parameter & Value & Source \\
\noalign{\smallskip}
\hline
\noalign{\smallskip}
  R.A.\; (Ep=2016.0) & 04$^\mathrm{h}$\,20$^\mathrm{m}$\,57.20$^\mathrm{s}$ & Gaia DR3 \\
  Dec.\; (Ep=2016.0) & $-$68$\degr$\,06$\arcmin$\,09.68$\arcsec$ & Gaia DR3 \\
  $B$ & $9.91\pm0.03$ & CDS Simbad \\
  $V$ & $9.38\pm0.02$ & CDS Simbad \\
  $B_T$ & $10.021\pm0.026$ & \citet{hog2000} \\
  $V_T$ & $9.438\pm0.021$ & \citet{hog2000} \\
  $J$ & $8.476\pm0.020$ & \citet{cutri2003} \\
  $H$ & $8.256\pm0.024$ & \citet{cutri2003} \\
  $Ks$ & $8.175\pm0.023$ & \citet{cutri2003} \\
  $G$ & $9.3063\pm0.0028$ & Gaia DR3 \\
  $G_{\mathrm{RP}}$ & $9.5625\pm0.0028$ & Gaia DR3 \\
  $G_{\mathrm{BP}}$ & $8.8875\pm0.0038$ & Gaia DR3 \\
  Parallax $\varpi$\; [mas] & $7.1364\pm0.0098$ & Gaia DR3\tablefootmark{(a)} \\
  Distance [pc] & $140.13 \pm 0.19$ & Gaia DR3\tablefootmark{(a)} \\
\noalign{\smallskip}
\hline
\noalign{\smallskip}
  \teff\ [K] & $6189\pm55$ & Spectroscopy \\\relax
  \feh\                   & $0.10\pm0.03$ & Spectroscopy \\\relax
  \logg\ (cgs)        & $4.25\pm0.07$ & Spectroscopy \\
  \vsini\ [\kms] &  $8.0\pm0.5$ & Spectroscopy \\ 
  $L_{\star} \ [L_{\odot}]$ & $2.71^{+0.16}_{-0.14}$ & $\varpi$ \& Photometry \\
  $R_{\star}$ [$R_{\odot}$] & $1.425_{-0.041}^{+0.048}$ & $\varpi$ \& Photometry\tablefootmark{(b)} \\
  $M_{\star}$ [$M_{\odot}$] & $1.227\pm0.052$ & Isochrones \\
  $t_{\star}$ [Gyr] & $3.2\pm0.9$ & Isochrones \\
  \logrhk & $-4.991\pm0.003$ & Spectroscopy \\
\noalign{\smallskip}
\hline                        
\end{tabular}
\tablefoot{Values in the bottom part of the table were derived in this work. \\ \tablefoottext{a}{Zero-point correction from \citet{lindegren2021} applied.} \\ \tablefoottext{b}{The role of \texttt{PARSEC} isochrones for the $R_{\star}$ estimate is to transform absolute magnitudes into luminosity via bolometric correction tables provided with the evolutionary grids \citep{marigo2017}.}}
\end{table}

\subsection{Spectroscopic parameters} 
\label{ssec: Stellar modelling}

We derived the spectroscopic parameters of TOI-282 from the co-added ESPRESSO spectrum, which has a S/N of $\sim$1400 per pixel at 550\,nm. (Sect.~\ref{ssec: ESPRESSO spectroscopic data}). The modelling of the spectrum was performed with the code Spectroscopy Made Easy\footnote{\url{https://www.stsci.edu/~valenti/sme.html}.} \citep[\texttt{SME}, version 5.2.2;][]{Valenti1996, Piskunov2017}, along with synthetic spectra generated using \texttt{ATLAS12} atmospheric models \citep{Kurucz2013}. \texttt{SME} automatically computes synthetic spectra and compares them with the observed spectrum to derive the spectroscopic parameters. We modelled one parameter at a time, utilising spectral features sensitive to different photospheric parameters, and iterating until convergence of all parameters was reached. We measured the projected rotational velocity \vsini\ by simultaneously fitting unblended iron lines. Throughout the modelling, we kept the macro- and micro-turbulent velocities (\vmac\ and \vmic) fixed to 4.1~\kms\ \citep{Doyle2014} and 1.2~\kms\ \citep{Bruntt2008}, respectively. An exhaustive description of the procedure is provided in \citet{Persson2018}. Our results are listed in Table~\ref{tab:star}.

As a sanity check, we used the code \texttt{specmatch-emp}\footnote{\url{https://github.com/samuelyeewl/specmatch-emp}.} \citep{Yee2017}, which derives the spectroscopic parameters by comparing observed spectra with a grid of nearly 400 high-resolution, high S/N template spectra acquired with the HIRES spectrograph. Prior to executing \texttt{specmatch-emp}, we lowered the resolution and adjusted the format of the ESPRESSO co-added spectrum to ensure compatibility with the HIRES template spectra. This analysis provides an effective temperature of \teff\,=\,5980\,$\pm$\,110\,K, in fairly good agreement with the \texttt{SME} results. We also employed the spectral energy distribution (SED) fitting software \texttt{astroARIADNE}\footnote{\url{https://github.com/jvines/astroARIADNE}.} \citep[][]{Vines2022} to measure the interstellar reddening along the line of sight to the star and derive a first estimate of the stellar radius. We used the spectroscopic parameters derived with \texttt{SME} and the broad-band magnitudes listed in Table~\ref{tab:star}, along with the Gaia DR3 offset-corrected parallax $\varpi$ \citep{GaiaDR3,lindegren2021}. We fitted the SED using three different atmospheric model grids from  \texttt{Phoenix\,v2} \citep{Husser2013}, \citet{Castelli2004}, and \citet{Kurucz1993}. The radius and visual extinction were calculated via Bayesian model averaging. We found that the star has a radius of $R_\star$\,=\,$1.46\pm 0.02$~$R_{\odot}$ and suffers a negligible reddening of $A_\mathrm{v}$\,=\,$0.06\pm0.04$.

\subsection{Stellar luminosity, mass, radius, and age} 
\label{ssec: Stellar parameters}

To refine the stellar radius $R_{\star}$ and derive the luminosity $L_{\star}$, mass $M_{\star}$, and age $t_{\star}$, we followed the multi-photometric approach described in \citet{bonfanti2025}. Briefly, for each broadband magnitude listed in Table~\ref{tab:star}, we first built an input set of parameters made of the \texttt{SME}-derived \teff, \met, and \logg, along with the Gaia DR3 offset-corrected parallax $\varpi$. We then interpolated pre-computed grids of \texttt{PARSEC}\footnote{\textsl{PA}dova and T\textsl{R}ieste \textsl{S}tellar \textsl{E}volutionary \textsl{C}ode; \url{https://stev.oapd.inaf.it/cgi-bin/cmd}.} v1.2S isochrones and tracks \citep{marigo2017} via the isochrone placement routine \citep{bonfanti2015,bonfanti2016} by using each input set as a constraint. We ended up with ten different sets of estimates for $L_{\star}$, $R_{\star}$, $M_{\star}$, and $t_{\star}$ (one for each input magnitude). After checking the mutual consistency of the results via the $\chi^2$-based criterion outlined in \citet{bonfanti2021}, we merged the respective distributions and determined $L_\star$\,=\,$2.71_{-0.14}^{+0.16}$~$L_\odot$, $R_{\star}=1.425_{-0.041}^{+0.048}~R_{\odot}$, $M_{\star}=1.227\pm0.052\,M_{\odot}$, and $t_{\star}=3.2\pm0.9$~Gyr (Table~\ref{tab:star}). The stellar radius is in agreement with that obtained using \texttt{astroARIADNE}, corroborating our results. Assuming that TOI-282 is seen equator-on ($i_\star$\,=\,90$^\circ$), the stellar radius and projected rotational velocity (Table~\ref{tab:star}) imply a rotation period of $P_{\mathrm{rot}, \star}\,=\,9.1\,\pm\,0.7$~d. Based on the $B-V\,=\,0.53\,\pm\,0.04$ colour index and the median \logrhk\,=\,$-$4.991\,$\pm$\,0.003 (Table~\ref{tab:star}), the rotation-activity empirical relations from \citet{Mamajek2008} and \citet{SuarezMascareno2016} provide a rotation period of 12\,$\pm$\,2~d and 13\,$\pm$\,5~d, respectively, which are consistent with the value we estimated from \vsini. 

\citet{Dransfield12022} derived a gyro-chronological age of $1.1\,\pm\,0.1$~Gyr for TOI-282, using the activity relation from \citet{Mamajek2008} and a \logrhk\,=\,$-4.56\pm0.05$ estimated from the GALEX FUV excess. We note that the latter value is significantly lower than our median \logrhk\,=\,$-$4.991\,$\pm$\,0.003, which is likely to be more accurate as it is directly extracted from the Ca\,{\sc ii}\,H\,\&\,K lines (Sect.~\ref{sec: HARPS high-resolution spectroscopy}). Based on our \logrhk\ measurement, the activity relation from \citet{Mamajek2008} yields a gyro-chronological age of $2.7\,\pm\,1.3$~Gyr, which agrees with our isochronal age of $3.2\,\pm\,0.9$~Gyr, suggesting that TOI-282 is likely older than previously reported.

\section{Methods} 
\label{sec: Methods}

\subsection{Radial velocity de-trending} 
\label{ssec: Radial velocity extraction and detrending}

The intrinsic magnetic activity of the host star might mimic, or even conceal, Doppler signals induced by orbiting planets \citep[see, e.g.][]{Haywood2014}. Therefore, modelling stellar activity is vital to disentangle planetary signals from the so-called 'stellar noise' \citep{Saar1997, Queloz2001, Boisse2011, Dumusque2018, Faria2020}. Activity indicators and CCF shape parameters can show a strong correlation with RV measurements if time series are affected by stellar activity \citep{Hatzes1996, Queloz2001, Boisse2011, Figueira2013}. Since the asymmetry of the absorption lines correlates with stellar activity, the CCF skewness $\gamma$ has proved to trace the RV variability induced by active regions carried around by stellar rotation \citep{Simola2019,Bonfanti2023}.

For each RV time series (HARPS, ESPRESSO 1st set, and ESPRESSO 2nd set), we used the CCF profile diagnostics and one activity indicator against which we detrended the RV measurements. Similarly to \citet{luque2023Nature} and \citet{fridlund2024}, the model adopted in our analysis is defined as follows
    \begin{equation} \label{eq: stellar activity model}
    \begin{aligned}[b]
        RV = \beta_0 & + \sum_{k=1}^{k_t}\beta_{k,t}t^k + \sum_{k=1}^{k_F}\beta_{k,F}\,\mathrm{FWHM}^k + \sum_{k=1}^{k_A}\beta_{k,A}A^k + \\
        & + \sum_{k=1}^{k_{\gamma}}\beta_{k,\gamma}\gamma^k + \sum_{k=1}^{k_{\Phi}}\beta_{k, \Phi}\Phi^k,
    \end{aligned}
    \end{equation}
where the parameters $\beta$ are the polynomial coefficients, while $k_t, k_F, k_A, k_{\gamma}$, and $k_{\Phi}$ are the orders of the polynomial functions fitted against time $t$, FWHM, contrast $A$, skewness $\gamma$, and one activity indicator $\Phi$ (e.g. $\log{R'_{\text{HK}}}$, H$_{\alpha}$, Na\,D1, Na\,D2), if present.

    \begin{table*}[!t]
        \centering
        \begin{threeparttable}
        \caption{Parameters of the TOI-282 planetary system.}
        \begin{tabular}{lp{3cm}p{3cm}p{3cm}}
            \hline
            \toprule
             Parameter & TOI-282\,b & TOI-282\,c & TOI-282\,d \\ [0.3ex]
             \hline 
             \midrule
             Orbital period $P$ \ $^\ddag$ (d) & \(22.891054^{+0.000085}_{-0.000090}\) &  \(55.99712^{+0.00034}_{-0.00037}\) &   \(84.2899^{+0.0018}_{-0.0015}\) \\ [0.3ex]
             Mid-time of reference transit \(T_0 \ ^\ddag \) & \(8344.8150^{+0.0039}_{-0.0036}\) &  \(8337.2807^{+0.0055}_{-0.0048}\) &   \(8355.5685^{+0.0075}_{-0.0086}\) \\ [-1ex]
             $(\text{BJD}_{\text{TBD}} -2\,450\,000)$ &&& \\ [0.3ex]
             Transit Depth d$F$ \ (ppm) & \(301^{+46}_{-41}\) &  \(715\pm17\) &  \(407\pm27\) \\ [0.3ex]
             Impact Parameter $b$  & \(0.927^{+0.026}_{-0.019}\) & \(0.31^{+0.09}_{-0.12}\) &  \(0.25^{+0.13}_{-0.15}\) \\ [0.3ex]
             Orbital Inclination i (°) & \(87.94^{+0.09}_{-0.10}\) & \(89.61^{+0.15}_{-0.12}\) & \(89.77^{+0.15}_{-0.13}\) \\ [0.3ex]
             Scaled semi-major axis \(a/R_{\star}\) & \(25.71^{+0.81}_{-0.89}\) & \(46.7\pm1.6\) & \(61.3^{+1.9}_{-2.1}\) \\ [0.3ex]
             Transit duration \(T_{14}\) \ (hours) & \(2.69^{+0.34}_{-0.29}\) & \(8.957\pm0.065\) &  \(10.32^{+0.26}_{-0.23}\) \\ [0.3ex]
             Radial velocity semi-amplitude $K$ \ \((\text{m s}^{-1}) \) & \(1.22\pm0.31\) &  \(1.35\pm0.28\) & $<$1.24\,$^\dag$ \\ [0.3ex]
             Planetary Radius \(R_p \ (R_{\oplus})\) & \(2.69\pm0.23\) &  \(4.13^{+0.16}_{-0.14}\) &   \(3.11\pm0.15\) \\ [0.3ex]
             Planetary Mass \(M_p \ (M_{\oplus})\) & \(6.2\pm1.6^\ast \) &  \(9.2\pm2.0^\ast\) &   \(5.8^{+0.9\ast \ast}_{-1.1} \) \\ [0.3ex]
             Planetary Density \( \rho_p \ (\text{g cm}^{-3})\) & \(1.8^{+0.7 \ast}_{-0.6}\) &  \(0.7\pm0.2^{\ast}\) &   \(1.1^{+0.3\ast \ast}_{-0.2} \) \\ [0.3ex]
             Equilibrium Temperature \(T_{eq}\) (K) & \(867\pm16\) & \(643\pm12\) & \(561 \pm 11\) \\ [0.3ex]
             \hline
             \hline
             \multicolumn{4}{c}{Other Parameters}   \\ 
             \hline
             LD coefficient \(u_1\) & \multicolumn{3}{c}{\(0.2426\pm0.0068\)} \\ [0.3ex]
             LD coefficient \(u_2\) & \multicolumn{3}{c}{\(0.3081\pm0.0043\)} \\ [0.3ex]
             RV Jitter HARPS (m s$^{-1}$) & \multicolumn{3}{c}{$0.848^{+0.055}_{-0.035}$} \\ [0.3ex]
             RV Jitter ESPRESSO 1st set (m s$^{-1}$) & \multicolumn{3}{c}{$1.411^{+0.076}_{-0.048}$} \\ [0.3ex]
             RV Jitter ESPRESSO 2nd set (m s$^{-1}$) & \multicolumn{3}{c}{$0.43^{+0.15}_{-0.10}$} \\ [0.3ex]
             \bottomrule
        \end{tabular}
        \begin{tablenotes}
          \small
          \item \(\ddag\): Derived assuming linear ephemerides. 
          \item \(\dag\): Three $\sigma$ upper limit. With $K_d$\,=\(0.28^{+0.32}_{-0.20}\)~\ms, the Doppler reflex motion induced by TOI-282\,d remains undetected in the RVs. 
          \item $\ast$: Derived from the RV+phot. joint modelling conducted with \texttt{MCMCI}. The dynamical analysis carried out with \texttt{TRADES} provides consistent results of $M_b= 6.7^{+1.7}_{-0.8}~M_\oplus$ and $M_c = 10.0^{+1.0}_{-2.0}~M_\oplus$ (Sect.~\ref{sec: Dynamical analysis}).
          \item $^{\ast \ast}$ Derived from dynamical analysis using \texttt{TRADES} (Sect.~\ref{sec: Dynamical analysis}).
        \end{tablenotes}
        \label{tab: nonlinear ephe}
        \end{threeparttable}
    \end{table*}

\subsection{Transit photometry and RV joint analysis} 
\label{ssec: Joint analysis}

To determine the orbital parameters of TOI-282\,b, c, and d and retrieve their masses, radii, and mean densities, we conducted a joint analysis of the 58 TESS transit light curves and 81 HARPS and ESPRESSO RV measurements. We investigated the possibility of splitting the RV time series into piecewise stationary segments using the breakpoint method \citep{Bai2003} as presented in \cite{Simola2022}, but found no breakpoint.

We analysed the data within a Markov chain Monte Carlo (MCMC) framework as implemented in the \texttt{MCMCI} software suite \citep{Bonfanti2020}. We imposed Gaussian priors on the stellar mass $M_{\star}$, radius $R_{\star}$, iron content \feh, and effective temperature \teff\ (Table~\ref{tab:star}). This approach was adopted to achieve two primary aims: drive the convergence of the transit fitting by imposing a prior on the stellar density $\rho_{\star}$ via the $M_{\star}$ and $R_{\star}$ Gaussian priors; retrieve the best-fitting quadratic limb darkening (LD) coefficients following interpolation within the LD's grid pre-computed with the \cite{Espinoza2015}'s code. For the modelling of the transit light curves, we re-parametrised the LD coefficients as described in \cite{holman2006}.

For the remaining model parameters (i.e. transit depths d$F\equiv\left(\frac{R_p}{R_{\star}}\right)^2$, impact parameters $b$, orbital periods $P$, reference transit times $T_0$, and RV semi-amplitudes $K$), we adopted uniform unbounded priors, except for those set by physical limits. We also explored the possibility of eccentric orbits by applying uniform priors to ($\sqrt{e}\cos{\omega}$, $\sqrt{e}\sin{\omega}$), with $e$ being the eccentricity and $\omega$ the argument of peri-centre. We performed two tests, either by constraining the eccentricities to $e \leq 0.3$, following the results of \citetalias{Dransfield12022}, or by leaving them entirely unconstrained within physical limits. We found that as the range of eccentricities allowed in the MCMC analysis increases, the parameter convergence worsens, indicating that the available data do not provide sufficient constraints on the planetary eccentricities. We also found that models with $e \neq 0$ are disfavoured based on the $\Delta$BIC criterion \citep[e.g.][]{Kass1995, Trotta2007} and adopted circular orbits for all three planets (BIC$_{e=0}=16240$ and BIC$_{e\neq0}=16310$).

The TESS transit light curves were de-trended against the ancillary vectors presented in Sect.~\ref{ssec: TESS photometry} using polynomial functions. For the RV time series, we followed the method described in Sect.~\ref{ssec: Radial velocity extraction and detrending}. Polynomial orders were determined using the BIC minimisation criterion through \texttt{MCMCI} mini-runs comprising 10\,000 steps. For the ESPRESSO 1st and 2nd data sets, we used only the three CCF profile diagnostics (FWHM, $A$, $\gamma$). For the HARPS data set, along with the respective FWHM, $A$, and $\gamma$, we also considered H$_{\alpha}$, as we found it to be the spectral activity indicator that most effectively lowers the BIC when coupled with the CCF shape parameters, and also the only one whose periodogram displays peaks at frequencies consistent with the stellar rotation frequency and its harmonics (see Appendix~\ref{sec: Frequency analysis} for more details). Most TESS transit light curves only require a normalisation scalar (0-order polynomial), likely because the PDCSAP pipeline has effectively removed most instrumental effects and the star does not show strong photometric variability. We also carried out an initial \texttt{MCMCI} run with 200\,000 steps to assess the contributions of white and red noise to the light curves and RVs, following the methodology described in \cite{Pont2006} and \cite{Bonfanti2020}. This allowed us to rescale the photometric errors and obtain reliable uncertainties for the derived parameters. For the final analysis, we opted for three independent runs of 200\,000 steps each (burn-in set to 20\% of the total number of steps) and checked the convergence via the Gelman-Rubin test \citep{Gelman1992}.

\subsection{Transit timing variations' analysis} 

\begin{figure}[!htbp]
        \centering        {\includegraphics[width=0.99\linewidth, trim={0.3cm, 0.4cm, 0.3cm, 0.2cm}, clip]{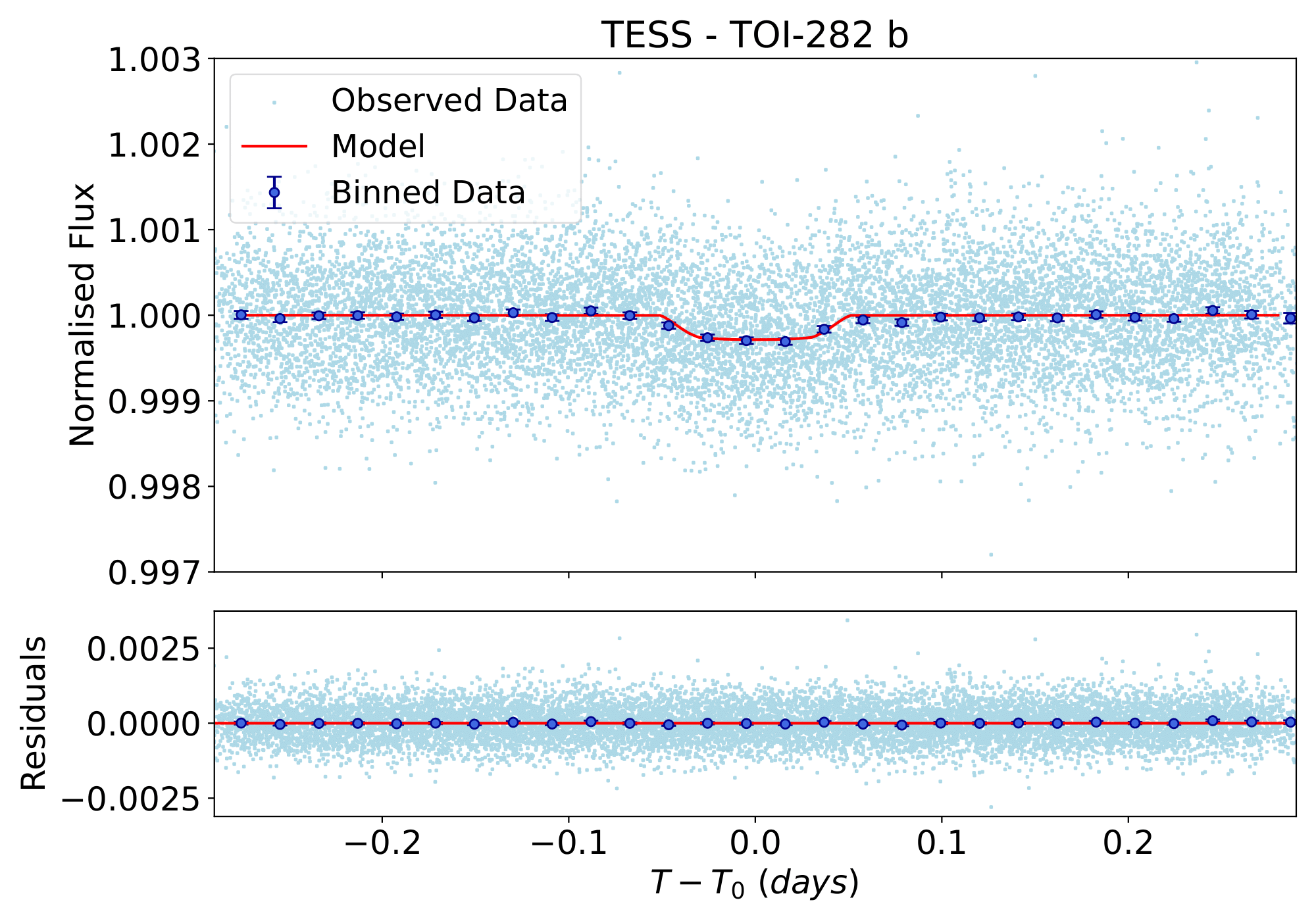}\label{fig: LC_b}} \\        {\includegraphics[width=0.99\linewidth, trim={0.3cm, 0.4cm, 0.3cm, 0.2cm}, clip]{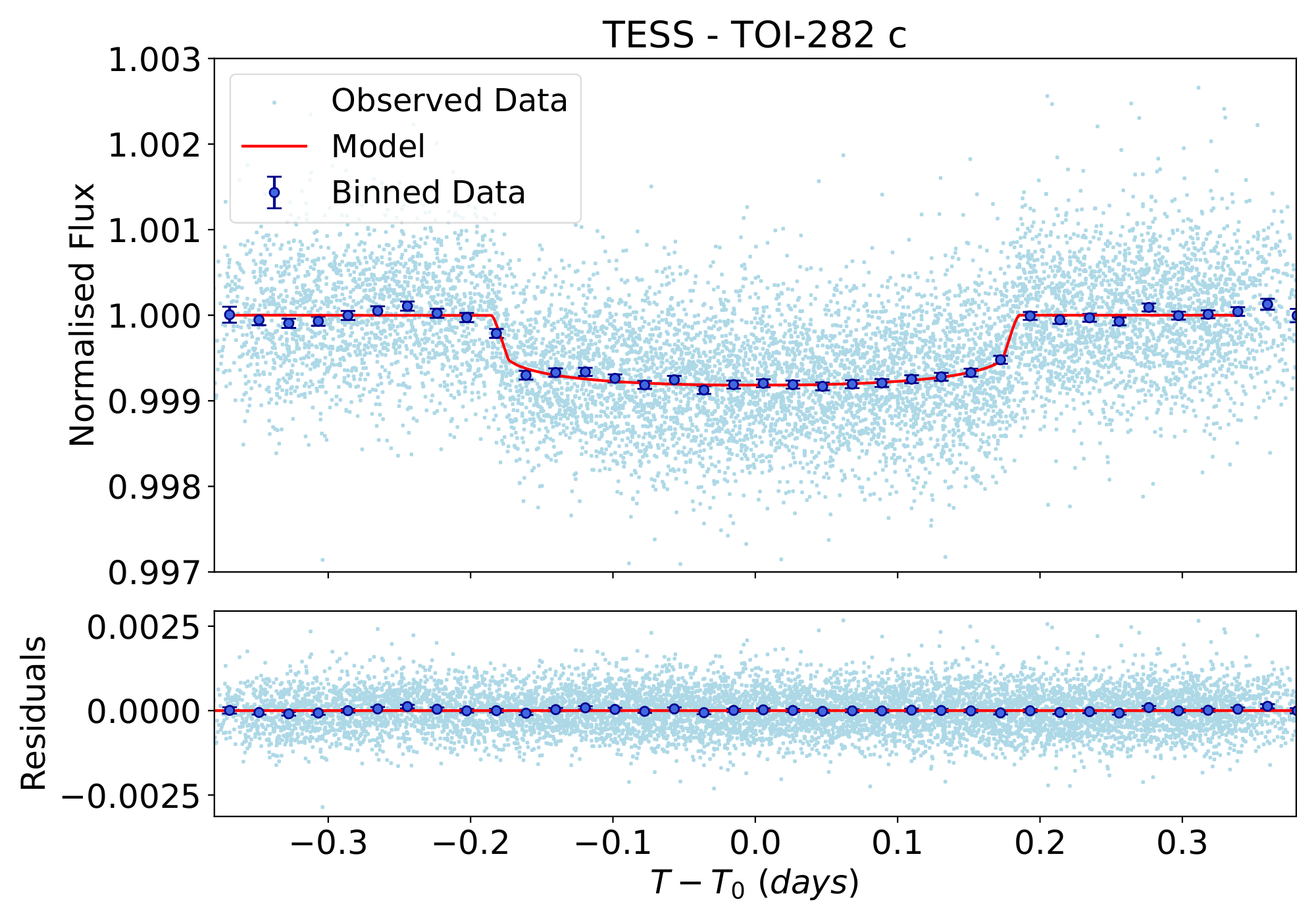}\label{fig: LC_c}} \\        {\includegraphics[width=0.99\linewidth, trim={0.3cm, 0.4cm, 0.3cm, 0.2cm}, clip] {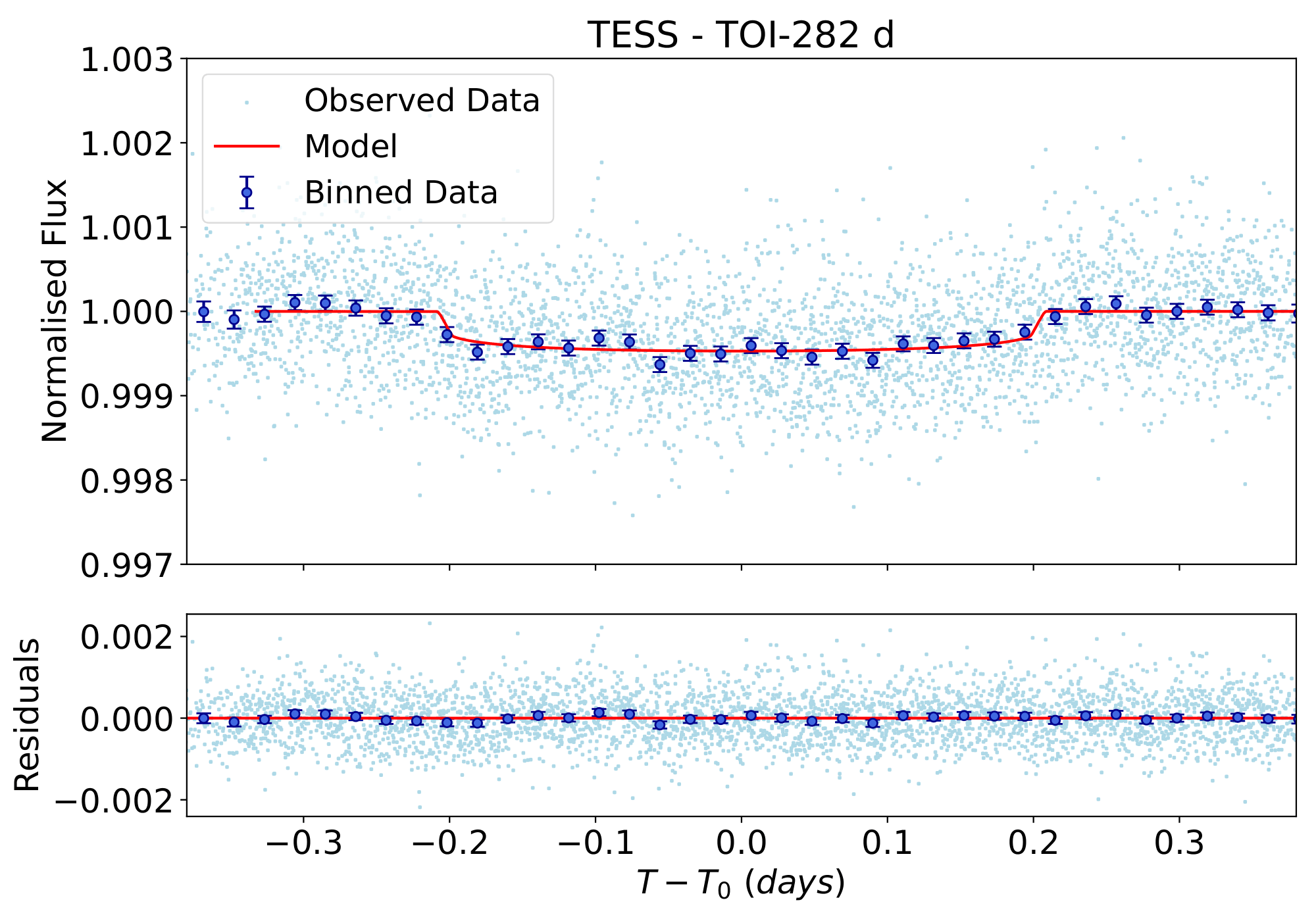}\label{fig: LC_d}} 
        \caption{\emph{From top to bottom}: folded TESS transit light curves of TOI-282\,b, c, and d. The photometric data points are plotted with light-blue circles. The dark blue circles mark the 30-minute binned data points. The best-fitting models are over-plotted with red curves. The lower panels display the residuals of the models.} \label{fig: LCs}
    \end{figure}

\begin{figure}[!htbp]
    \centering
    {\includegraphics[width=0.99\linewidth, trim={0.3cm, 0.4cm, 0.3cm, 0.2cm}, clip]{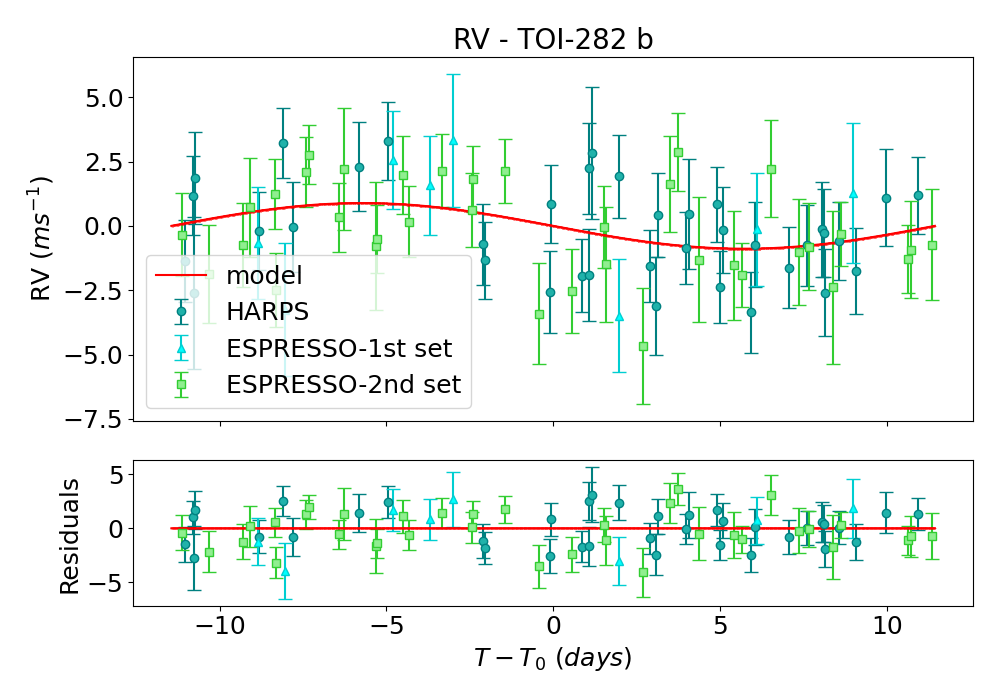}\label{fig: RV_b}} \\
        {\includegraphics[width=0.99\linewidth,trim={0.3cm, 0.4cm, 0.3cm, 0.2cm}, clip]{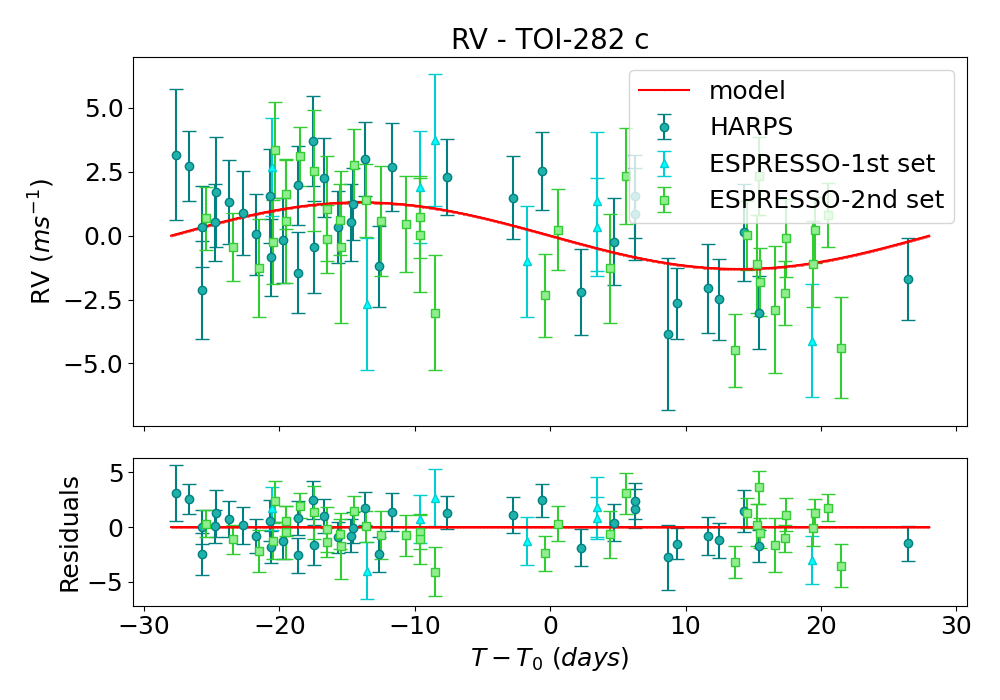}\label{fig: RV_c}} \\
        {\includegraphics[width=0.99\linewidth, trim={0.3cm, 0.4cm, 0.3cm, 0.2cm}, clip]{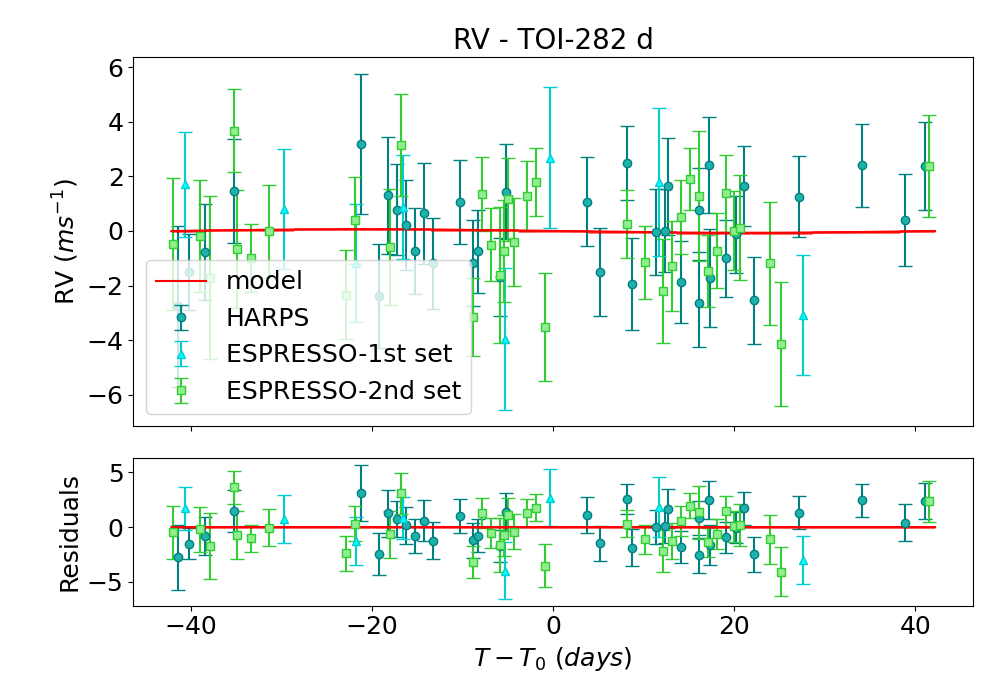}\label{fig: RV_d}}
        \caption{\emph{From top to bottom}: folded RV stellar signals induced by TOI-282\,b, c, and d, and best-fitting RV models (red curve). The data includes 37 HARPS spectra (blue circular points) and 8 + 36 ESPRESSO spectra (cyan triangular and green square points, respectively). The lower panels show the residuals to the models.} \label{fig: RVs}
    \end{figure}

To retrieve the orbital period ($P_i$) and mid-time of reference transit ($T_{0,\,i}$) of each planet, we first modelled the transit LCs and HARPS and ESPRESSO RVs, assuming linear ephemerides (i.e. constant orbital periods). As pointed out by \citetalias{Dransfield12022}, the orbital periods of TOI-282\,c and d are very close to the 3:2 commensurability, which likely induces strong gravitational interactions between the two planets, causing the periodic variations of their orbital periods.
To measure the TTVs, we repeated the \texttt{MCMCI} analysis, allowing the transit timings to deviate from the linear ephemerides predictions. Briefly, for each planet $i$, we fixed $P_{i}$ and $T_{i,\,0}$ to the values derived assuming linear ephemerides (Table~\ref{tab: nonlinear ephe}), while allowing the other transit parameters to vary, including the mid-time ($T_{i,\,j}$) of each transit event $j$. The chains of the model parameters converged according to the Gelman-Rubin statistic, except for 6 of the 33 available transits of TOI-282 b, where the LCs exhibit an excess of noise, making the detection of this nearly grazing (b\,$\approx$\,0.9; Table~\ref{tab: nonlinear ephe}) shallow ($\sim$300~ppm) transit challenging. A similar issue occurred with one transit LC of TOI-282~d, which only covers the ingress of the event. The medians of the posterior distributions of the system parameters, along with the 68.3\% confidence intervals, are listed in Table~\ref{tab: nonlinear ephe}, while Figs.~\ref{fig: posteriorlinb}, \ref{fig: posteriorlinc}, and \ref{fig: posteriorlind} in Appendix~\ref{sec: Appendix} show the posterior distributions of the model parameters. The transit LCs of the three planets, after accounting for TTVs, are shown in Fig.~\ref{fig: LCs}, while the RV curves are displayed in Fig.~\ref{fig: RVs}.

For each planet $i$ and transit $j$, we calculated the TTV$_{i,\,j}$ defined as the difference between the observed and predicted mid-times, i.e. TTV$_{i,\,j}$\,=\,$T_{i,\,j} - (T_{i,\,0} + n_{i,\,j}\,P_{i})$, where $n_{i,\,j}$ is the transit epoch. We performed a linear fit to the observed mid-times $T_{i,\,j}$ as a function of the transit epoch $n_{i,\,j}$, and recomputed the orbital period ($P_{i}$) and the mid-time of the reference transit ($T_{i,\,0}$). We evaluated the statistical significance of the TTV$_i$ for each planet through the reduced--$\chi^2_i$ ($\overline{\chi}^2_i$) defined as
\begin{equation}
    \centering
    \overline{\chi}^2_i = \frac{1}{N_i-2} \sum_{j=0}^{N_i-1} \left( \frac{\mathrm{TTV}_{i,\,j}}{\sigma_{i,\,j}} \right)^2,\nonumber
\end{equation}
where $N_i$ is the number of transits available for each planet $i$, $N_i$\,$-$\,2 are the degrees of freedom, and $\sigma_{i,\,j}$ are the uncertainties of the observed mid-times $T_{i,\,j}$. 
We obtained $\overline{\chi}^2_{b}$\,=\,0.37, $\overline{\chi}^2_{c}$\,=\,93.1, and $\overline{\chi}^2_{d}$\,=\,16.7, which implies significant TTVs for TOI-282\,c and d (Fig.~\ref{fig:O-Cs}), with the transit mid-times being on average $\sim$9$\sigma$ and $\sim$3.5$\sigma$ away from the linear ephemerides predictions.

\section{Transit light curve and radial velocity data analysis results} \label{sec: LC and RV data analysis results}

The planetary parameters are consistent with those reported by \citetalias{Dransfield12022}, except the transit depth (d$F_b$) and the impact parameter ($b_b$) of TOI-282\,b. \citetalias{Dransfield12022} found a depth of $\mathrm{d}F_{b,\mathrm{D22}}=208^{+9}_{-11}$~ppm and an impact parameter of $b_{b,\mathrm{D22}}=0.8007^{+0.0055}_{-0.0084}$, whereas our estimates are d$F_b$\,=\,\(301^{+46}_{-41}\)~ppm and $b_b$\,=\,\( 0.927^{+0.026}_{-0.019} \). Their results were also obtained under the assumption of linear ephemerides, using uniform priors on \(R_b/R_{\star}\), \( (R_b + R_\star)/a \), and \(\cos{i_b}\). Looking at Table~3 of \citetalias{Dransfield12022}, we note that their median values are very close to the bounds of the uniform priors, suggesting that their posterior distributions might be truncated and, consequently, their estimates and associated uncertainties might be biased. We suspect that the uniform priors adopted by \citetalias{Dransfield12022} are too narrow on a few occasions. This holds especially for the priors of the model parameters that are used to retrieve the impact parameter \(b_b\) of planet~b. For instance, the priors adopted by \citetalias{Dransfield12022} for \( (R_b + R_\star)/a \) and \(\cos{i_b}\) are \(\mathcal{U}(0.05,0.06)\) and \(\mathcal{U}(0.00,0.04)\), respectively, and their estimates for these model parameters are \( (R_b + R_\star)/a_\mathrm{D22} = 0.05017^{+0.00022}_{-0.00012}\) and \(\cos{i}_{b,\mathrm{D22}}=0.03964^{+0.00024}_{-0.00040}\). The transit depth d\(F_b\) is also quite affected, since the planet-to-star radius ratio found by \citetalias{Dransfield12022} for TOI-282~b of \(R_p/R_{\star,\mathrm{D22}}=0.0144^{+0.00031}_{-0.00037}\) is less than 2$\sigma$ away from the upper limit of the corresponding uniform prior of \(\mathcal{U}(0.008,0.015)\) adopted by the authors. Furthermore, the phase-folded transit light curve of TOI-282\,b, as presented in Fig.~7 of \citetalias{Dransfield12022}, exhibits a transit of $\sim$300~ppm, which agrees with our estimate (see the upper panel of Fig.~\ref{fig: LCs} and Table~\ref{tab: nonlinear ephe}) but it does not match the value reported by the authors in their Table~3. We also identified what we believe are two oversights in Table~3 of \citetalias{Dransfield12022} and on the NASA Exoplanet Archive\footnote{\url{https://exoplanetarchive.ipac.caltech.edu}.}. Specifically, the mid-time of reference transit for TOI-282\,c ($T_{0,c}$) does not lie within its uniform prior. This parameter, along with the full transit duration of TOI-282\,b ($T_{14,b}$), does not match the plots shown in Figs.~2 and 7 of \citetalias{Dransfield12022}.

The $\sim$7-year baseline of the TESS photometry allowed us to retrieve a more precise set of ephemerides for the three transiting planets than those presented by \citetalias{Dransfield12022}, which are based on $\sim$2.8 years of TESS observations. We improved the uncertainties in the reference transit times ($T_0$) and orbital periods $P$ by a factor of $\sim$1.8 and $\sim$4, respectively (Table~\ref{tab: nonlinear ephe}). We found that TOI-282\,b, c, and d have radii of $R_b = 2.69 \pm 0.23 \ R_{\oplus}$, $R_c = 4.13^{+0.16}_{-0.14} \ R_{\oplus}$, and $R_d = 3.11 \pm 0.15 \ R_{\oplus}$, respectively. The radii of the two outermost planets agree with those obtained by \citetalias{Dransfield12022}. On the other hand, there is a $2\sigma$ tension for the innermost planet's radius, which very likely arises from the small prior range for $\text{d}F_b$ adopted by \citetalias{Dransfield12022}.

We found that the two innermost transiting planets TOI-282\,b and TOI-282\,c have masses of $M_b$\,=\,6.2\,$\pm$\,1.6~$M_{\oplus}$ (26\% relative precision) and $M_c$\,=\,9.2\,$\pm$\,2.0~$M_{\oplus}$ (22\% relative precision), respectively. To our knowledge, these are the first RV mass measurements of the two planets with a significance greater than 3$\sigma$. Using a small subset of the HARPS and ESPRESSO RVs presented in this paper\footnote{Seven HARPS and 8 ESPRESSO measurements.}, \citetalias{Dransfield12022} derived a mass of $18.5^{+9.1}_{-7.6}$~$M_\oplus$ for TOI-282\,b, which agrees with our mass determination within $\sim$1.6$\sigma$. Unfortunately, the Doppler reflex motion induced by the outermost transiting planet TOI-282\,d remains undetected in our Doppler measurements. We found an RV semi-amplitude of $K_d\,=\,0.28_{-0.20}^{+0.32}$~\ms, which implies a 3$\sigma$ mass upper limit of $M_d \le 9.7~M_{\oplus}$. Nonetheless, we performed a dynamical analysis of the system and measured the mass of planet d, as described later in Sect.~\ref{sec: Dynamical analysis}. The radii and masses of TOI-282\,b and c translate into bulk densities of $\rho_b=1.8_{-0.6}^{+0.7}$~g\,cm$^{-3}$ and $\rho_c$\,=\,0.7\,$\pm$\,0.2$~$g\,cm$^{-3}$, with relative precisions of \(36\%\) and \(28\%\), respectively. 

\begin{figure}[!t]
    \centering
    \includegraphics[width=0.85\linewidth, trim={0.25cm, 0.28cm, 0.23cm, 0.25cm}, clip]{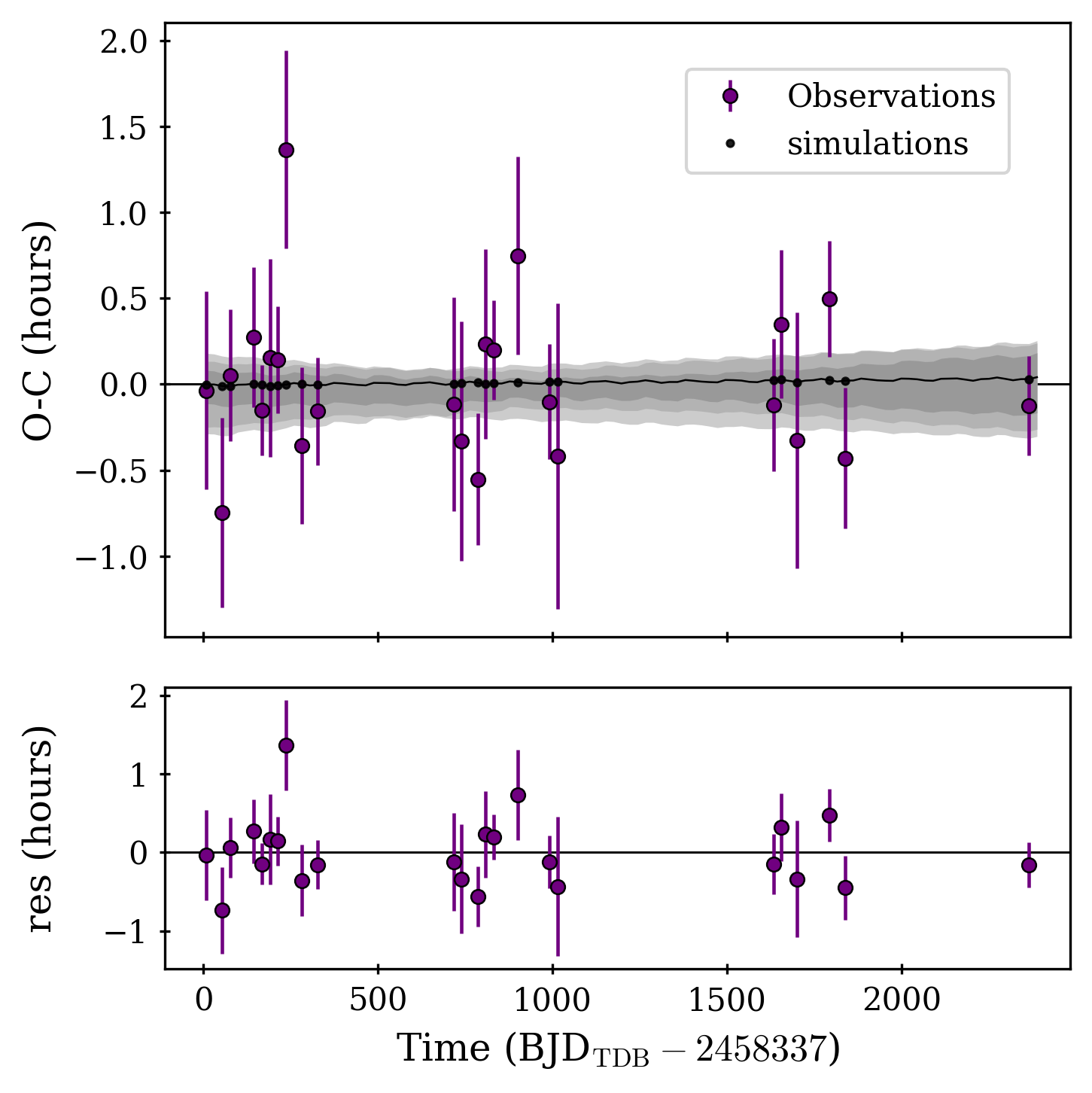}
    \caption{Observed minus calculated (O$-$C) TTV diagram for TOI-282\,b, showing TTV as a function of time. The shaded regions represent the $1, 2,~\mathrm{and} ~3\sigma$ confidence interval derived from the posterior distributions of the model parameters of our dynamical analysis.}
    \label{fig:OC_b}
\end{figure}

\begin{figure}[!htbp]
        \centering
        \includegraphics[width=0.85\linewidth, trim={0.25cm, 0.28cm, 0.23cm, 0.25cm}, clip]{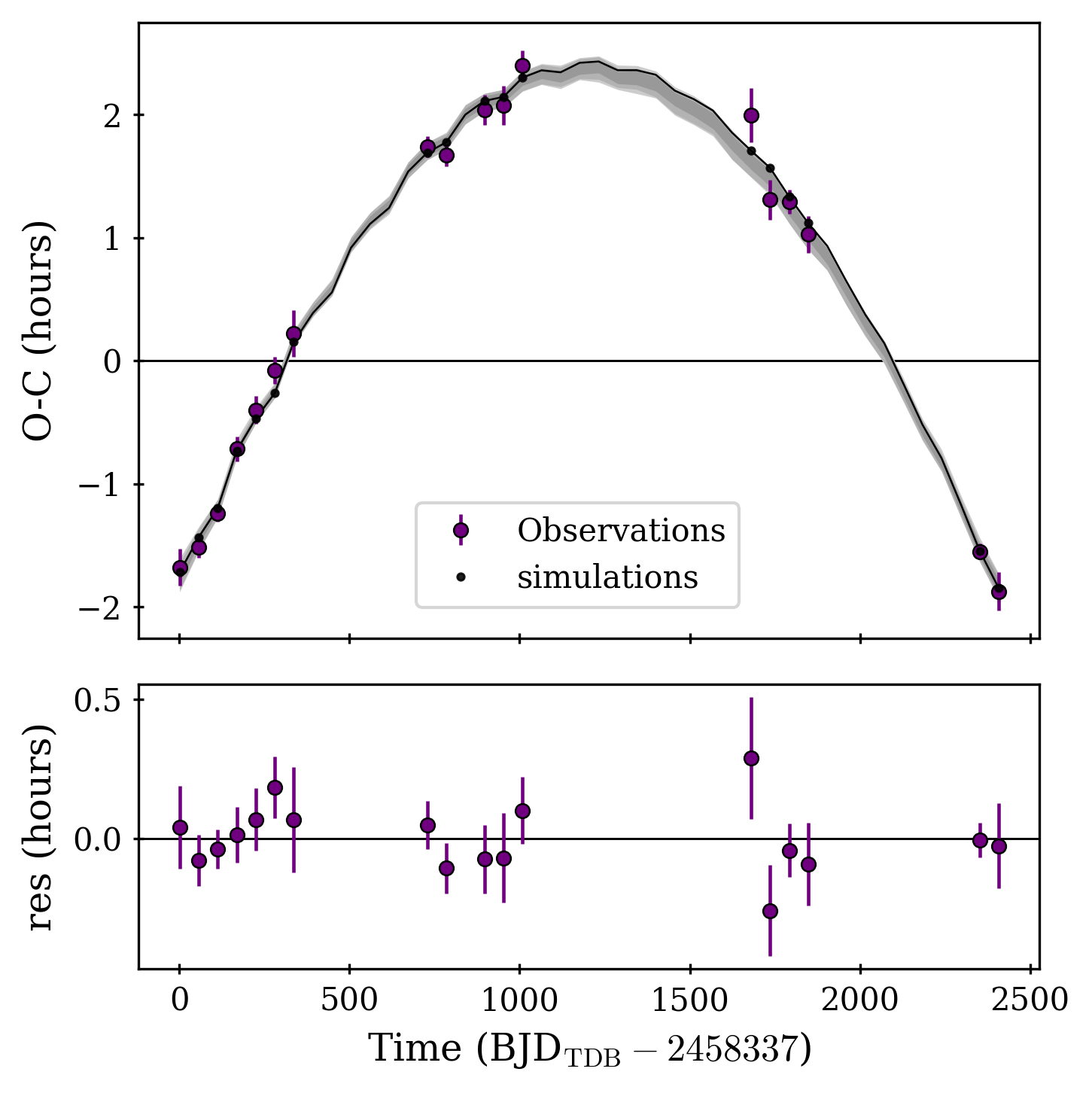}
        \includegraphics[width=0.85\linewidth, trim={0.25cm 0.28cm 0.23cm 0.25cm}, clip]{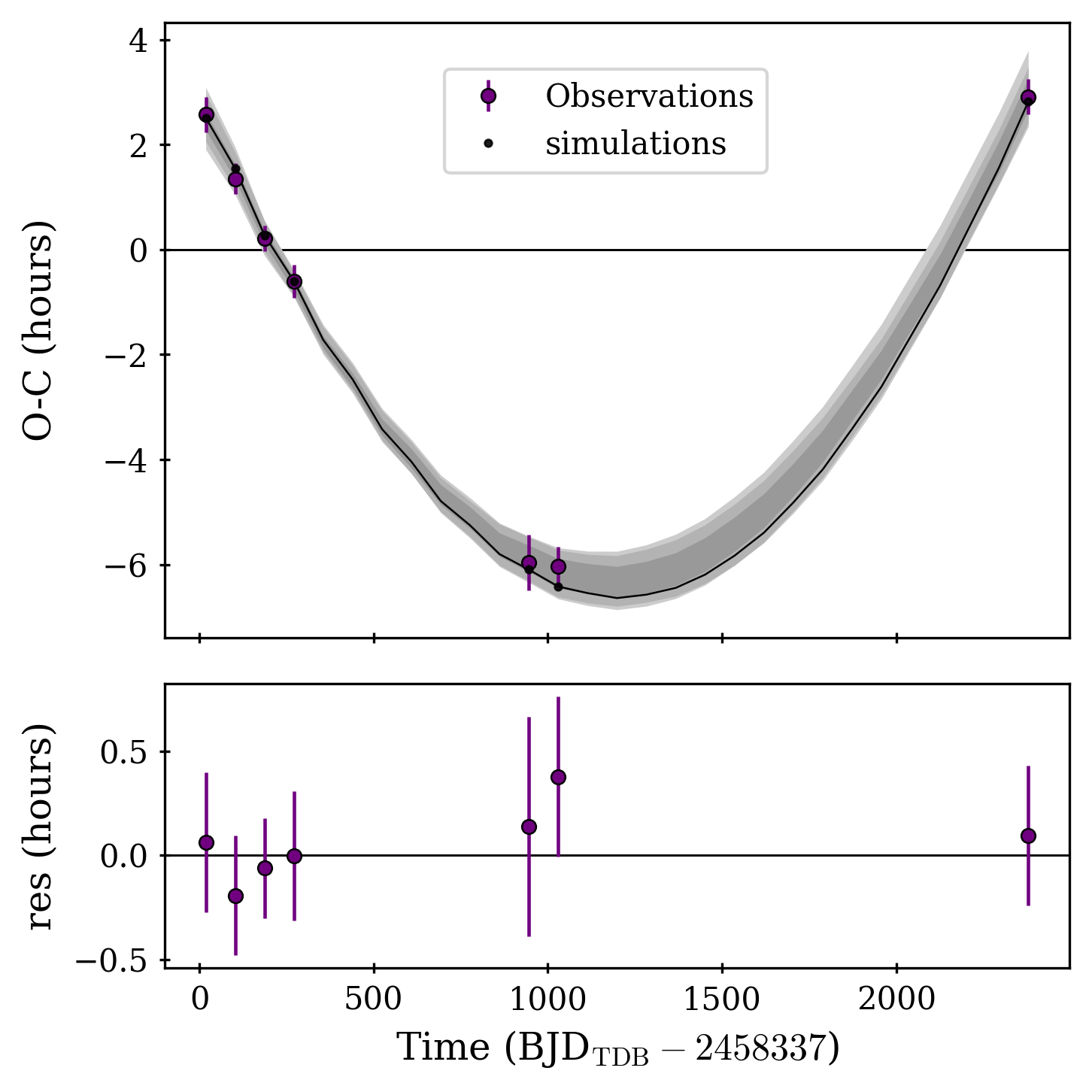}
    \caption{Observed minus calculated (O$-$C) TTV diagrams for TOI-282\,c (upper panel) and TOI-282\,d (lower panel), showing TTV as a function of time. The shaded regions represent the $1, 2,~\mathrm{and} ~3\sigma$ confidence interval derived from the posterior distributions of the model parameters of our dynamical analysis. Both planets exhibit anti-correlated TTV exceeding the transit time uncertainties.}
    \label{fig:O-Cs}
\end{figure}

\section{Dynamical modelling} \label{sec: Dynamical analysis}
\subsection{Transit timing variation and RV dynamical analysis}

In multi-planet systems, mutual gravitational interactions between the planets lead to small deviations from Keplerian motions, producing observable variations in the timing of transits \citep{Agol2005, holman_2005}. By tracking these deviations from a linear ephemeris, known as transit timing variations (TTVs), we can probe the system's dynamics, detect non-transiting companions, and constrain planetary masses and orbital parameters. Furthermore, by combining RV measurements with TTVs, we can break the intrinsic degeneracies of each method, enabling a more complete and robust characterisation of the system.

We found that TOI-282\,b displays no significant TTVs (Fig.~\ref{fig:OC_b}), in agreement with the findings of \citetalias{Dransfield12022}. TOI-282\,c and d show strong TTVs with a peak-to-peak amplitude of $\sim$4 and $\sim$8~hours, respectively (Fig.~\ref{fig:O-Cs}). Although the Doppler reflex motion induced by planet d remains undetected in our RV time series, its TTVs anti-correlate with those of TOI-282\,c (Fig.~\ref{fig:O-Cs}), suggesting a possible resonant configuration with planet c. To measure the mass of TOI-282\,d, we performed a N-body integration dynamical analysis of the planetary system, simultaneously modelling TTVs (Tables~\ref{table:T0_b}, \ref{table:T0_c}, and \ref{table:T0_d}) along with the de-trended RVs (Sect.~\ref{sec: Methods}, and Tables~\ref{table:HARPS_RV_Table} and \ref{table:ESPRESSO_RV_Table}). For TOI-282\,b, we excluded 9 transit timings, as their uncertainties $\sigma_T \gtrsim  4$\,hours. The analysis was conducted using TRAnsits and Dynamics of Exoplanetary Systems \citep[\texttt{TRADES},][]{Borsato2014, Borsato2019, Borsato2021}\footnote{\url{https://github.com/lucaborsato/trades}.}, a \texttt{Fortran-Python} hybrid code that simultaneously fits transit timings and RV measurements, while self-consistently integrating planetary orbits. 
To cover the observational baseline, we performed a dynamical integration spanning $2410$ days, initialised at the reference epoch $2\,458\,337~\mathrm{BJD_{TDB}}$.

For each planet we fitted the orbital period $P$, the planet-to-star mass ratios $M_p/M_{\star}$, the eccentricities $e$, the argument of peri-centre of the planetary orbit $\omega_p$, and mean longitude~$\lambda$\footnote{Defined as $\lambda = \mathcal{M} + \omega_p + \Omega$, where $\mathcal{M}$ is the mean anomaly, $\omega_p$ the argument of peri-centre, and $\Omega$ the ascending node longitude.}. To mitigate biases, we adopted the $\left(\sqrt{e} \cos \omega_p, \sqrt{e} \sin \omega_p\right)$ parametrisation, rather than modelling $e$ and $\omega_p$ separately. Planetary radii $R_\mathrm{p}$, stellar radius $R_\star$ and mass $M_\star$, and orbital inclinations $i$ were fixed to the values listed in Tables~\ref{tab:star} and \ref{tab: nonlinear ephe}. The longitude of the ascending node was fixed to $\Omega = 180^\circ$ for all planets.  
We adopted uniform priors for all the fitted parameters (Table~\ref{table:fit_TRADES_orbital_parameters}). We also fitted an RV jitter term ($\sigma_{\text{jitter}}$), by adopting $\log_2 \sigma_{\text{jitter}}$ as a step parameter.

The initial exploration of the parameter space was conducted via the \pyde\footnote{\url{https://github.com/hpparvi/PyDE}.} differential evolution algorithm \citep{Storn_97, Parviainen_2016}, using a population size of 120 for 60\,000 iterations. The solutions were then used as initial parameters for \emcee{} \citep{foremanmackey_2013}, where 120 walkers explored the parameter space for 400\,000 steps, approximately six times the number of free parameters in our model. Following \cite{Nascimbeni2024}, our \emcee{} sampler combined the differential evolution proposal (80\% walkers; \citealt{nelson_2014}) with the snooker variant (20\% walkers; \citealt{terbraak_2008}). After confirming chain convergence via Geweke \citep{geweke_1991}, Gelman-Rubin \citep{gelmanrubin_1992}, autocorrelation \citep{goodman_weare_2010}, and visual diagnostics, we discarded 100\,000 steps as burn-in and applied a thinning factor of 100. Parameter uncertainties were quantified using the 68.3\% highest density intervals (HDIs) of the marginalised posteriors. Inferred values correspond to the maximum a posteriori (MAP) estimates derived from the posterior distributions.

The parameter estimates and their uncertainties, along with the adopted priors, are listed in Table~\ref{table:fit_TRADES_orbital_parameters}, whereas the observed minus calculated (O$-$C) diagrams for TOI-282\,b, c, and d are shown in Figs.~\ref{fig:OC_b} and \ref{fig:O-Cs}. The anti-correlation patterns follow the model prediction made by \citetalias{Dransfield12022}. Our dynamical analysis allowed us to retrieve the orbital configuration of TOI-282\,b,\,c, and d. This approach provides the masses of the two innermost planets $M_b =6.7^{+1.7}_{-0.8}$ and $M_c = 10^{+1}_{-2}$, which are in very good agreement with the \texttt{MCMCI}-derived ones (Table~\ref{tab: nonlinear ephe}). Moreover, we obtained a robust mass determination for TOI-282\,d of $\mathrm{M_{d} = 5.8_{-1.1}^{+0.9}}~\mathrm{M_{\oplus}}$ ($\sim$15-20\% relative precision). When combined with the radius of $R_d = 3.11 \pm 0.15 \ R_{\oplus}$, the mass of TOI-282\,d implies a mean density of $\mathrm{\rho_{d} = 1.1_{-0.2}^{+0.3}}~\mathrm{g~cm^{-3}}$ ($\sim25$\% relative precision). We acknowledge that, while the available TTV and RV data provide a relatively precise mass determination, the O$-$C diagrams (Fig.~\ref{fig:O-Cs}) partially cover the $\sim$9\,000-day TTV super-period \citep[Eq.~5 from][]{Lithwick2012} of the near 3:2 pair. To accurately measure the planetary mass via TTV, a better sampling of the entire super-period is needed to reduce possible systematic uncertainties and degeneracies between eccentricity and mass solutions.

\begin{figure}[!t]
        \centering
        \subfloat[][]{\includegraphics[width=0.99\linewidth]{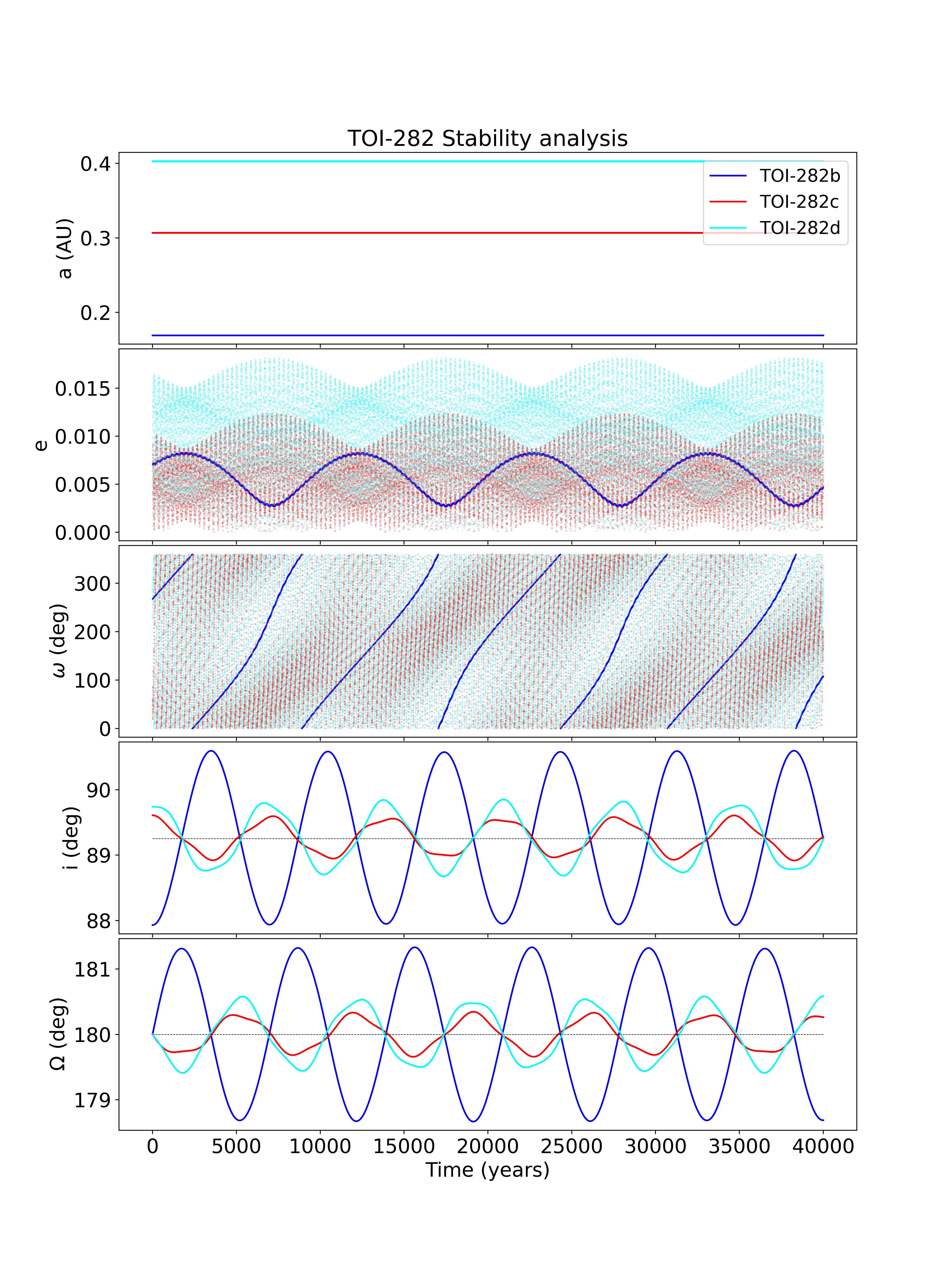}\label{fig: oscillations}} \\
        \subfloat[][]{ \includegraphics[width=0.99\linewidth]{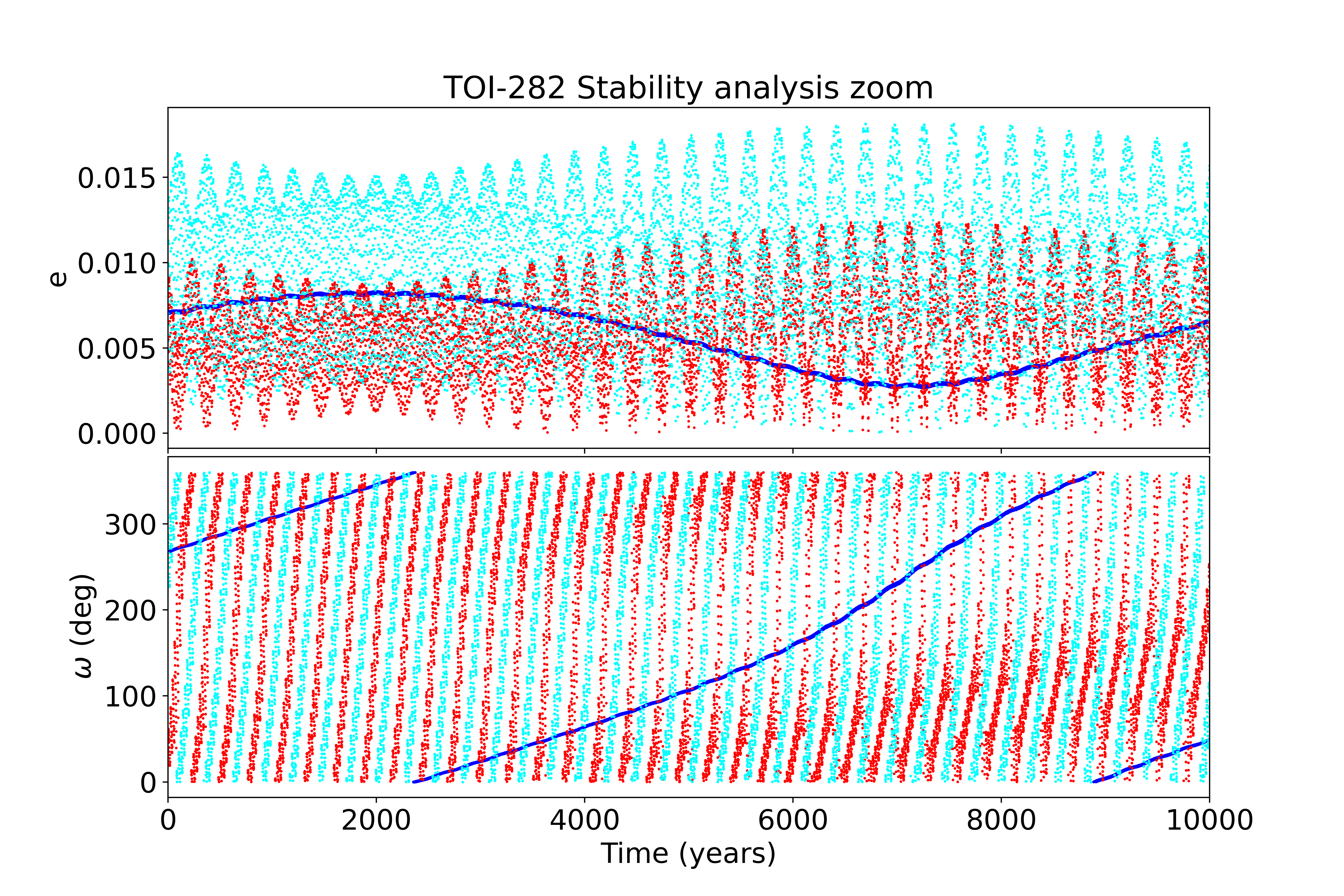}\label{fig: oscillations_zoom}} \\
        \caption{Simulated evolution of the orbital parameters over $10^5$ years. Figure b shows a zoom over the first $10^4$ years on the eccentricity and argument of peri-centre plots. The colours blue, red, cyan refer respectively to planets TOI-282\,b, TOI-282\,c, TOI-282\,d.
        }\label{fig: stability}
\end{figure}

\subsection{Stability analysis}
As an additional check in our dynamical analysis, \trades{} assesses the system’s Hill stability using the AMD-Hill criterion (Eq.~26; \citealt{petit_2018}), which is based on the angular momentum deficit (\texttt{AMD}; \citealt{laskar_1997, laskar_2000, laskar_petit_2017}). We find that the entire posterior distribution satisfies the AMD-Hill stability criterion, indicating long-term dynamical stability. To further evaluate the stability and potential chaotic behaviour of the posterior solutions, accounting for planet-planet interactions, mean-motion resonances, and possible planetary ejections, we performed N-body integrations with \texttt{rebound} \citep{rein_liu_2012, rein_tamayo_2016}. Specifically, we employed the mean exponential growth factor of nearby orbits (\texttt{MEGNO}; $\langle \mathrm{Y} \rangle$) chaos indicator \citep{cincotta_simo_2000, cincotta_2003}, as implemented in \texttt{rebound}. We defined a dynamically stable planetary configuration as satisfying $\langle \mathrm{Y} \rangle \leq 2.5$ \citep{cincotta_simo_2000}, with planetary ejection occurring if any body’s semi-major axis exceeds $100\,a_d$. Using the symplectic integrator \texttt{WHFast} \citep{rein_tamayo_2016}, we propagated orbits for 100~kyr with a time-step equal to 10\% of planet b’s orbital period (${t_{\rm step} \approx 0.1\,P_b}$). For MAP solution within the highest density interval (HDI), we found $\langle \mathrm{Y} \rangle < 2.5$, indicating a strongly stable dynamics. To assess broader posterior solutions, we ran the integration for 500 randomly selected posteriors. We found that $\sim$92\% of the simulations were stable.
    
We also simulated the evolution of the system for $10^5$~years, adopting a fourth-order Runge-Kutta (\texttt{RK4}) algorithm \citep{Runge1895}. For each planet, we studied the evolution of the semi-major axis $a$, eccentricity $e$, argument of peri-centre $\omega_p$, inclination $i$, and longitude of ascending node $\Omega$. Our results are shown in Fig.~\ref{fig: stability} for two different time intervals, with Fig.~\ref{fig: oscillations} showing the $10^5$-year evolution and Fig.~\ref{fig: oscillations_zoom} covering the first $10^4$~years. We found that:

\begin{itemize}
       \item The semi-major axes ($a$) do not change significantly over time, with relative variations never exceeding 0.1\% (Fig.~\ref{fig: oscillations}, first panel).
        
        \item The eccentricities ($e$) of TOI-282\,c and TOI-282\,d oscillate with a period of about $280$~years, reaching a maximum value of $\sim$0.01 (Fig.~\ref{fig: oscillations}, second panel, and Fig.~\ref{fig: oscillations_zoom}, first panel). The oscillations have a super-period that matches the TOI-282\,b eccentricity evolution period of about $10500$~years.
        
        \item A peri-centre precession period of $\sim$7150~years for TOI-282\,b, and $\sim$270~years for both planets c and d (Fig.~\ref{fig: oscillations}, third panel, and Fig.~\ref{fig: oscillations_zoom}, second panel). Along with the precession motion, we also found that the peri-centre angle of TOI-282\,b ($\omega_b$) oscillates with a semi-amplitude of 30~degrees in quadrature with the evolution of its eccentricity ($e_b$).
        
        \item The orbital planes display a precession motion around their mean value, with the inclinations ($i$) and corresponding longitudes of the ascending node ($\Omega$) evolving in quadrature (Fig.~\ref{fig: oscillations}, fourth and fifth panels). The primary period of about $6890$~years is given by the relative (anti-phased) oscillation between the angular momentum of TOI-282\,b and the total angular momentum of TOI-282\,c and TOI-282\,d. The orbital planes of the two resonant outer planets also oscillate in anti-phase around their average value with a period of about $1960$~years and a semi-amplitude of $\sim$0.1$^{\circ}$.
\end{itemize}

\subsection{Resonant angles evolution}
\label{section:mmr}

\begin{figure}[]
    \centering
    \includegraphics[width=0.49\textwidth, trim={0.32cm 0.25cm 0.5 0}, clip]{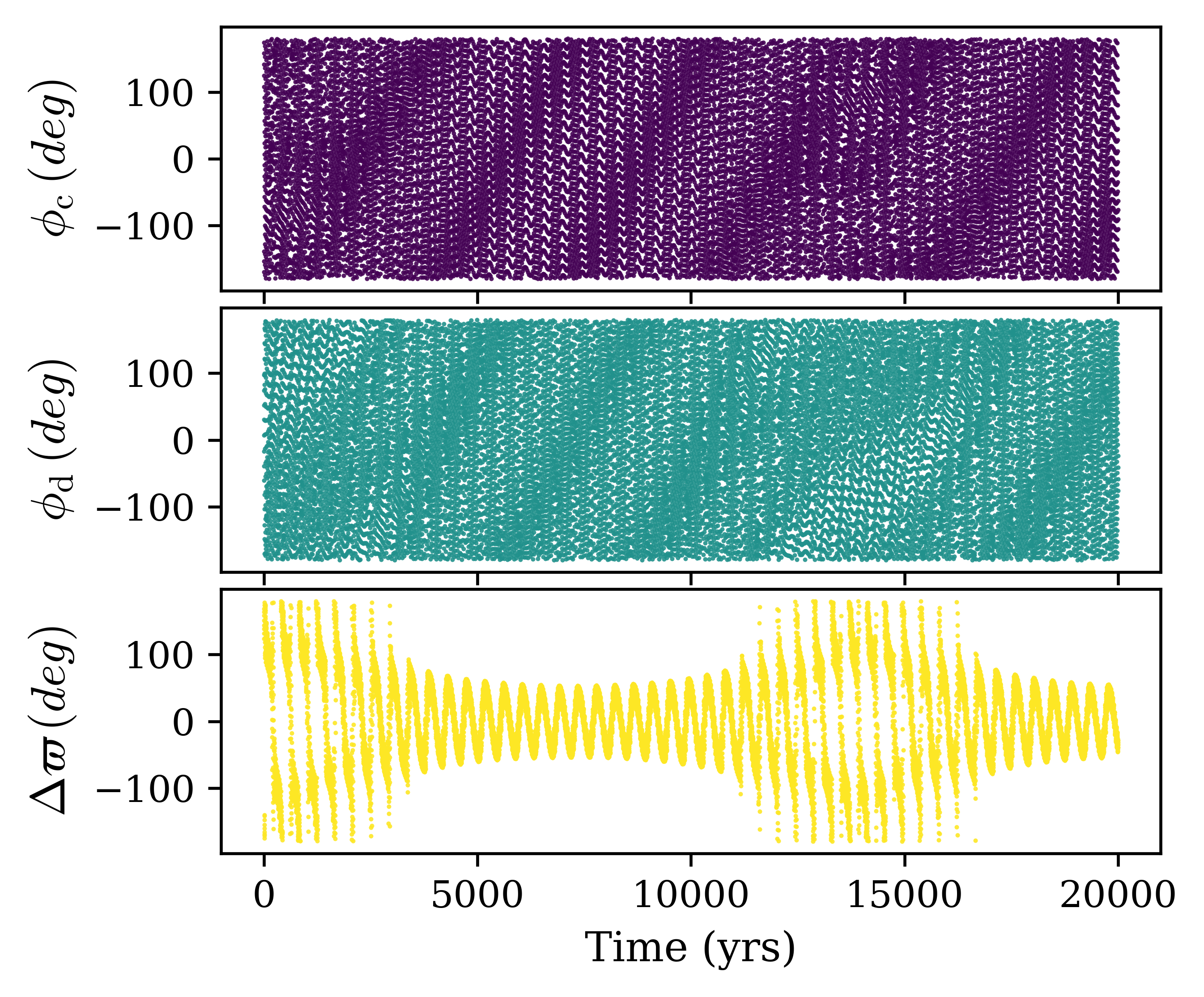}
    \caption{Resonant angle evolution for TOI-282\,c and d, for the MAP solutions. The two upper panels show the evolution of the critical angles for the 3:2 resonance ($\phi_c$ and $\phi_d$); the lower panel shows the difference between the longitudes of pericentre ($\Delta\varpi$).}
    \label{fig:Critical_angles}
\end{figure}

While period ratios near low-integer fractions may suggest proximity to a mean motion resonance, they do not definitively confirm it.
Following \citet{Lithwick2012}, we computed the so-called 'resonance offset', i.e. the deviation from the exact commensurability defined as $\Delta\,=\,(P_{out}/P_{in}) / (q/p)  -1\,$, where $q$\,=\,3 and $p$\,=\,2 for a 3:2 configuration. 
The resonance offset for TOI-282\,c and d is $\Delta$\,$\approx$\,0.0035, which agrees to the values found by N-body and hydrodynamical simulations \citep[e.g][]{silburt_rein_2015, ramos_2017}, and may indicate that the planets are in a resonance configuration, with librating resonant 'arguments'.
To uniquely assess if two outer planets are actually in an MMR, one must examine whether the critical resonant angles librate (oscillate around a fixed value) or circulate (continuously increase/decrease). Exact two-body first-order resonance requires libration within $\pm 180^\circ$ of one or both of: $\phi_{1} = q \lambda_{\rm 1} - (q + 1) \lambda_{\rm 2} + \varpi_{1}$, $\phi_{2} = q \lambda_{1} - (q + 1)\lambda_{2} + \varpi_{2}$. To investigate the resonant state of TOI-282\,c and TOI-282\,d, we integrated the MAP solutions using the N-body code \texttt{rebound} \citep{rein_liu_2012} and the symplectic integrator \texttt{WHFast} \citep{rein_tamayo_2016}, for a total of 20\,000 years. We computed the evolution of the resonant angles, defined as:
\begin{equation}
\begin{split}
    \phi_{c} &= 2\lambda_c - 3\lambda_d + \varpi_c, \\
    \phi_{d} &= 2\lambda_c - 3\lambda_d + \varpi_d,
\end{split}
\end{equation}
where $\varpi = \omega + \Omega$ denotes the longitude of peri-centre. As shown in Figure~\ref{fig:Critical_angles}, which also includes $\Delta\varpi \equiv \varpi_c - \varpi_d$, both angles circulate, indicating the system is not locked in a stable 3:2 resonance.

\section{The planetary system TOI-282} \label{sec: The planetary system TOI-282}

\begin{figure*}[!th]
        \sidecaption
        \centering
        \includegraphics[width=12cm, trim={1.8cm 0.3cm 2.4cm 1.7cm}, clip]{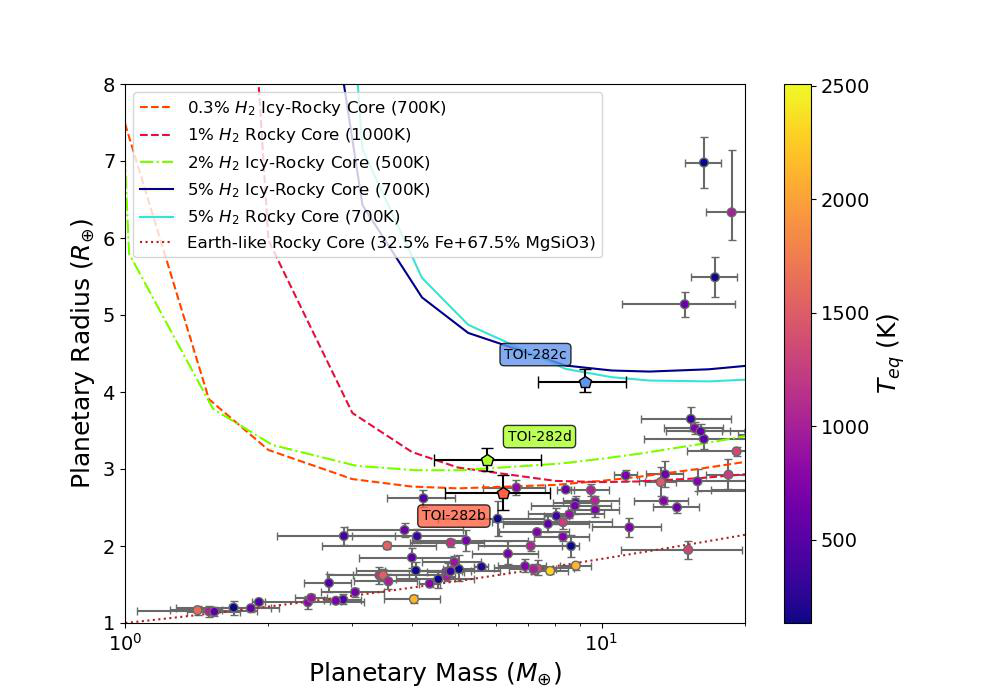}
        \caption{Mass-radius diagram showing the position of TOI-282\,b ($\text{T}_{\text{eq}} = 867 \pm 16~K$), TOI-282\,c ($\text{T}_{\text{eq}} = 643 \pm 12~K$), and TOI-282\,d ($\text{T}_{\text{eq}} = 561 \pm 11~K$)  with respect to the transiting planets whose masses and radii are known with a precision better than 30\% and 10\%, respectively (circles).  
        The coloured lines show the theoretical models from \cite{LiZeng2019}.}\label{fig: MR}
\end{figure*}

In agreement with \citetalias{Dransfield12022}, our results confirm that the F8 dwarf star TOI-282 hosts two sub-Neptunes (TOI-282\,b and d) and one Neptune-sized planet (TOI-282\,c), adopting the notation proposed by \citet{Kopparapu2018}. With P\,$<$\,100~days and period ratios $P_c/P_b$ and $P_d/P_c\,<\,5$, TOI-282~b, c, and d can be considered 'inner' planets with no significant gaps in the orbital periods \citep{Howe2025}. Following \cite{Mishra2023}, given their masses and radii, the three planets can be defined as 'similar'. For instance, their masses imply a coefficient of similarity of $C_S(M)$\,=\,0.008 and a coefficient of variation of $C_V(M)$\,=\,0.2, with the conditions of 'similarity' for TOI-282 being $|C_S(M)|$\,$\le$\,0.2 and $C_V(M)$\,$\le$\,0.7 \citep{Mishra2023}. 
This allows us to define the architecture of TOI-282 as a close-in "peas-in-a-pod" configuration, the most common architecture among the known multi-planet systems, representing $\sim$87\% of the peas-in-a-pod systems and $\sim$80\% of all the known planetary systems with at least 3 inner planets \citep{Howe2025}. These systems offer valuable insights into the formation and evolution of planets, being primary products of planet formation. The eccentricities and mutual inclinations of the planets are usually close to zero, and their internal compositions are similar. This suggests that compact, close-in systems represent the lowest energy state accessible to a forming planetary system. The three planets orbiting TOI-282 might have formed beyond the ice line and then migrated inward, reaching their current configuration through energy dissipation mechanisms. The similarity between the three planets would be a consequence of the same formation mechanism \citep{Weiss2022}.

We compared the positions of TOI-282~b, c, and d in the mass-radius (MR) diagram with those of other known exoplanets. We retrieved from the NASA Exoplanet Archive\footnote{\url{https://exoplanetarchive.ipac.caltech.edu/}.} masses and radii for the spectroscopically confirmed transiting exoplanets as of June 2025, and selected only those planets whose mass and radius were determined with relative precisions better than 30\% and 10\%, respectively. We chose the most recent reference for planets with different mass and radius estimates. Figure~\ref{fig: MR} shows the MR diagram of the planets that meet our criteria, colour-coded by their equilibrium temperature when available. To gain insight into the internal compositions of the planets, we also include the theoretical internal models from \cite{LiZeng2019}. 

TOI-282~b falls above a 'crowded' region of the MR diagram, in a quite degenerate parameter space, where it is difficult to obtain meaningful information on the composition of the planets. The planet could have an Earth-like rocky core ($32.5 \% \ \text{Fe} + 67.5\% \ \text{MgSiO}_3$), representing $99\%$ of its mass, surrounded by a $1\% \ \text{H}_2$ envelope\footnote{Assuming a 1~mbar surface pressure level; this assumption was made for all the models.}. Alternatively, TOI-282~b could be composed of a 1:1 mixture of H$_2$O ice and rocks, making it a new member of the so-called population of 'water worlds' \citep{Leger2004, Rogers2015, Mousis2020, Dorn2021, Luque2022}, with $\sim$0.3\% molecular hydrogen in its atmospheric envelope. From the composition of TOI-282~b, it would be possible to gain insight into its evolution and formation. Water worlds are believed to have formed beyond the ice line and then migrated towards the inner regions of the system \citep{Raymond2018, Bitsch2019, Izidoro2022}. On the other hand, Earth-like rocky planets are expected to be formed in situ, in the inner regions of the protoplanetary disc, and, in the absence of photo-evaporation processes, would retain their primordial atmosphere without being substantially enriched in volatiles \citep{Ikoma2012, Lee2014, Lee2016}. Due to the degeneracy, it is worth stressing that the superimposed theoretical models are not intended to give a definitive prediction on the composition of the planet, but rather a preliminary assessment of its possible internal structure \citep[e.g.][]{Rogers2010, Lopez2014, Dorn2015, Dorn2017, Otegi2020}.

With its low mean density ($\rho_c$\,=\,0.7\,$\pm$\,0.2~$\text{g\,cm}^{-3} $), TOI-282~c appears to be isolated in the MR diagram (Fig.~\ref{fig: MR}). Once again, theoretical models are found to be degenerate in this region of the parameter space. In fact, TOI-282~c might either have an Earth-like rocky core or might be composed of a 1:1 mixture of ice and rocks, with 5\% molecular hydrogen in its atmosphere. The outer transiting planet, TOI-282~d, also seems to be relatively isolated in the MR diagram (Fig.~\ref{fig: MR}). 
According to \cite{LiZeng2019}, it may also be a water world with $2\%$ molecular hydrogen in its envelope. If the three planets around TOI-282 were water worlds, they would follow the recent findings of \citet{Venturini2024}, where planets orbiting stars in the mass range $0.1 \le M_{\star} \le 1.5 \ M_{\odot}$ tend to be larger water worlds, rather than smaller rocky planets. 

The low mean density of TOI-282\,c and d, along with their relatively isolated positions in the MR diagram, could be linked to their nearly resonant state \citep[see Section~\ref{section:mmr};][]{Leleu2024}. Following \cite{Lee2016}, sub-Neptunes can form further out in the protoplanetary disc (in gas-poor regions), and enter an MMR state because of the subsequent inward migration. Alternatively, according to the "breaking the chains" model \citep{Bean2021}, sub-Neptunes in (close) MMR state may form in resonant chains due to the migration of planets in the protoplanetary discs and the positive torque at the inner edge of the disc \citep[e.g.][]{Goldreich1979, Weidenschilling1985, Masset2006, Terquem2007}. Instead, instabilities in the chains might lead to impacts, which may eject parts of the planet's primordial atmosphere \citep{Biersteker2019} and result in denser non-resonant objects, which could explain the nature of TOI-282\,b. The coplanarity of TOI-282\,c and d ($\Delta i_{c,d} = 0.2 \pm0.2^\circ$) strengthens this possibility, as planets in resonant chains tend to be coplanar \citep{Agol2021, Leleu2021}. At the same time, planet-planet scatterings are expected to increase the mutual inclinations ($\Delta i_{b,c} = 1.7 \pm0.2^\circ$  and $\Delta i_{b,d} = 1.8 \pm0.2^\circ$). 
  
The three planets in the TOI-282 system span a relatively small range of radii (2.7\,-\,4.1~$R_{\oplus}$) and masses (6.2\,-\,9.2~$M_{\oplus}$). They are all located above the so-called radius valley -- a dearth of small planets around $\sim$1.5\,-\,2.0~$R_{\oplus}$ \citep[see Fig.~4 in][]{Owen2017} -- and their relatively large radii can be explained by the presence of a thick atmosphere. This holds particularly for the two outermost transiting planets, as planets with radii $R$\,$\gtrsim$\,3~$R_{\oplus}$ are expected to have H/He in their atmosphere \citep{Jin2018}, and their long orbital periods ($P$\,$>$\,50~days) make a major mass loss due to photo-evaporation unlikely \citep{Gupta2020}. To explore the feasibility of atmospheric characterisation of the system, we determined for each planet the transmission and emission spectroscopy metrics \citep[TSMs and ESMs;][]{Kempton2018}. Starting from the closest planet to the star and moving further out, the TSMs are $\sim$39, 71, and 42, respectively, while the ESMs are $\sim$1.6, 1.8, and 0.7. The ESM is proportional to the expected S/N of a JWST planet occultation detection. We derived emission metrics values which are well below the recommended threshold ($\text{ESM}$\,=\,7.5), suggesting that it will probably not be possible to detect the occultations of the planets, mainly due to their low equilibrium temperatures (Table~\ref{tab: nonlinear ephe}). On the other hand, even though TOI-282~b, c, and d do not seem to be high-priority targets for transmission spectroscopy\footnote{The suggested threshold value is $\text{TSM}$\,=\,90 \citep{Kempton2018}.}, the bright host star in combination with the planets' radii, masses, and equilibrium temperatures led to acceptable metrics for atmospheric characterisation. Particularly, the three planets have a TSM higher than the lowest value of a top candidate in the radius bin of interest presented in \cite{Izidoro2022}. For instance, when compared to $\sim$1500 transiting exoplanets with known masses, we derived a quantile of 0.56, 0.39, and 0.54, respectively, making their atmospheric characterisation still feasible.

The TOI-282 system may be amenable to atmospheric characterisation via infrared high-resolution spectroscopy (HRS) with the ESO Extremely Large Telescope (ELT). Using $T_{eq}$ in Table~\ref{tab: nonlinear ephe} (zero albedo, full recirculation), the signal corresponding to 5 scale heights -- typical for strong molecular lines in the near infrared -- ranges from 33 to 60~ppm. Given the relatively bright host star (Table~\ref{tab:star}), with the ANDESatELT spectrograph \citep{Marconi2024}, it might be possible to detect TOI-282~c and d with one full transit (8\,-\,10 hours), while planet b would require 2\,-\,3 transits. Such observations would reveal the detailed chemical composition of the exoplanets, including potential N-bearing species associated with photochemical processes and the overall planet metallicity, the latter useful to break the degeneracy between interiors and envelopes.

With HRS, the ESM is not necessarily a good emission metric as the signal is driven by line contrast rather than continuum contrast. Using continuum contrast as a best-case scenario, ANDES would not be able to detect any of the planets easily. Even for TOI-282\,b, the flux ratio ranges between $10^{-10}$ at 1~$\mu$m and $10^{-7}$ at 1.8~$\mu$m, and it is 2\,-\,3 orders of magnitude worse for the cooler planets c and d. Such contrast ratios are beyond reach in spatially unresolved HRS. 

Nevertheless, in the METISatELT bandpass \citep[3\,-\,5~$\mu$m;][]{Brandl2021}, planet b would have a photospheric contrast of a few parts per million per line, which is about a factor of 10 smaller than typical 3$\sigma$ contrasts routinely achieved by current photon-limited HRS on 8\,-\,10-metre class telescopes. Considering the gain in collective area from the ELT, it may be possible to detect TOI-282\,b in 10\,-\,20 hours on sky, achievable in 2\,-\,3 scheduled nights of observations since HRS can observe both before and after a secondary eclipse.

\section{Conclusions} 
\label{sec: Conclusions}

TOI-282 is a late F-type bright (V=9.38) dwarf star, which hosts 3 long-period transiting small planets, with the outer two being close to a 3:2 MMR. The system was discovered and validated by \citetalias{Dransfield12022}, combining TESS space-based photometry with ground-based transit observations. We analysed 38 TESS sectors (15 more than those presented in \citetalias{Dransfield12022}), increasing the time baseline by nearly 3 years. We found that TOI-282~b, c, and d have radii of $R_b = 2.69 \pm 0.23 \ R_{\oplus}$, $R_c = 4.13^{+0.16}_{-0.14} \ R_{\oplus}$, and $R_d = 3.11 \pm 0.15 \ R_{\oplus}$, with a relative precision of $9\%$, $4\%$, and $5\%$, respectively. 

We performed an intensive RV follow-up of the star, securing 37 HARPS and 44 ESPRESSO high-precision Doppler measurements, which we jointly model with the TESS transit LCs.
We obtained a solid detection of the Doppler reflex motion induced by the two innermost planets ($\sim$4 and 5$\sigma$ level) and measured a mass of $M_b = 6.2 \pm 1.6 \ M_{\oplus}$ and $M_c = 9.2 \pm 2.0 \ M_{\oplus}$ for TOI-282~b and c, respectively. Although the outer planet (TOI-282~d) remains undetected in our RV time series, we determined its mass by performing a dynamical analysis of the system, leveraging the significant anti-correlated TTVs exhibited by TOI-282~c and TOI-282~d. 
We found that TOI-282\,d has a mass of $M_d=5.8^{+0.9}_{-1.1} \ M_{\oplus}$ ($\sim$15-20\% relative precision). The three planets have mean densities of  $\rho_b=1.8^{+0.7}_{-0.6} \ \text{g cm}^{-3}$, $\rho_c=0.7 \pm 0.2 \ \text{g cm}^{-3}$, and $\mathrm{\rho_{d} = 1.1_{-0.2}^{+0.3}}~\mathrm{g~cm^{-3}}$, which implies a relative precision of $\sim$36\%, 29\%, and 25\%, respectively. 

We also investigated the long-term dynamical evolution of the system to study the behaviour of the resonant angles associated with the two outer planets and the stability of the system's orbital architecture. According to our simulations,  while the system remains dynamically stable, the outer planet pair lies near, but does not maintain, a 3:2 MMR. We then placed the planets in an MR diagram to gain insight into their internal composition. By comparing the positions of the three planets on the MR diagram with internal structure models by \cite{LiZeng2019}, we found that TOI-282\,b and c might either be water worlds or have a rocky composition, while TOI-282\,d could be a water world. 

Further investigation of the system might break the internal composition degeneracy and provide insights into the nature and evolution of the three small planets transiting TOI-282. Future transmission spectroscopic observations with JWST and ELT are, in theory, feasible, as suggested by the derived metrics for atmospheric signals, and TOI-282 could be a high-priority target. TESS will observe the system from sector 93 (July 2025) through sector 98 (January 2026), providing an improved coverage of the TTV super-period. 

\section{Data availability}
Tables \ref{table:HARPS_RV_Table}, \ref{table:HARPS_ActivityIndicators_Table}, and \ref{table:ESPRESSO_RV_Table} are only available in electronic form at the CDS via anonymous ftp to \url{cdsarc.u-strasbg.fr} (130.79.128.5) or via \url{http://cdsweb.u-strasbg.fr/cgi-bin/qcat?J/A+A/}.

\begin{acknowledgements}

We thank the referee for the comments and suggestions that helped us to improve the quality of our paper.
We express our deepest gratitude to the ESO staff members for their unique support during the observations.
A.Ba and D.Ga. gratefully acknowledge the funding from the Physics Department of the University of Turin.
We thank Domenico Nardiello, David Rapetti, and Jon Jenkins for their precious efforts in reprocessing the TESS light curves to recover the transit photometry that falls within the gaps of the PDCSAP time series.   
We acknowledge financial support from the Agencia Estatal de Investigaci\'on of the Ministerio de Ciencia e Innovaci\'on MCIN/AEI/10.13039/501100011033 and the ERDF “A way of making Europe” through project PID2021-125627OB-C32, and from the Centre of Excellence “Severo Ochoa” award to the Instituto de Astrofisica de Canarias. 
P.Le. acknowledges that this publication was produced while attending the PhD program in Space Science and Technology at the University of Trento, Cycle XXXVIII, with the support of a scholarship co-financed by the Ministerial Decree no. 351 of 9th April 2022, based on the NRRP - funded by the European Union - NextGenerationEU - Mission 4 ``Education and Research'', Component 2 ``From Research to Business'', Investment 3.3 -- CUP E63C22001340001.

\end{acknowledgements}

\bibliographystyle{aa}
\bibliography{Bibliography} 

\begin{appendix}
\clearpage

\section{Frequency analysis of the HARPS and ESPRESSO time series} \label{sec: Frequency analysis}

\begin{figure*}[]
        \centering
        \includegraphics[width=0.8\linewidth, trim={2.6cm, 1.3cm, 3.75cm, 2.25cm}, clip]{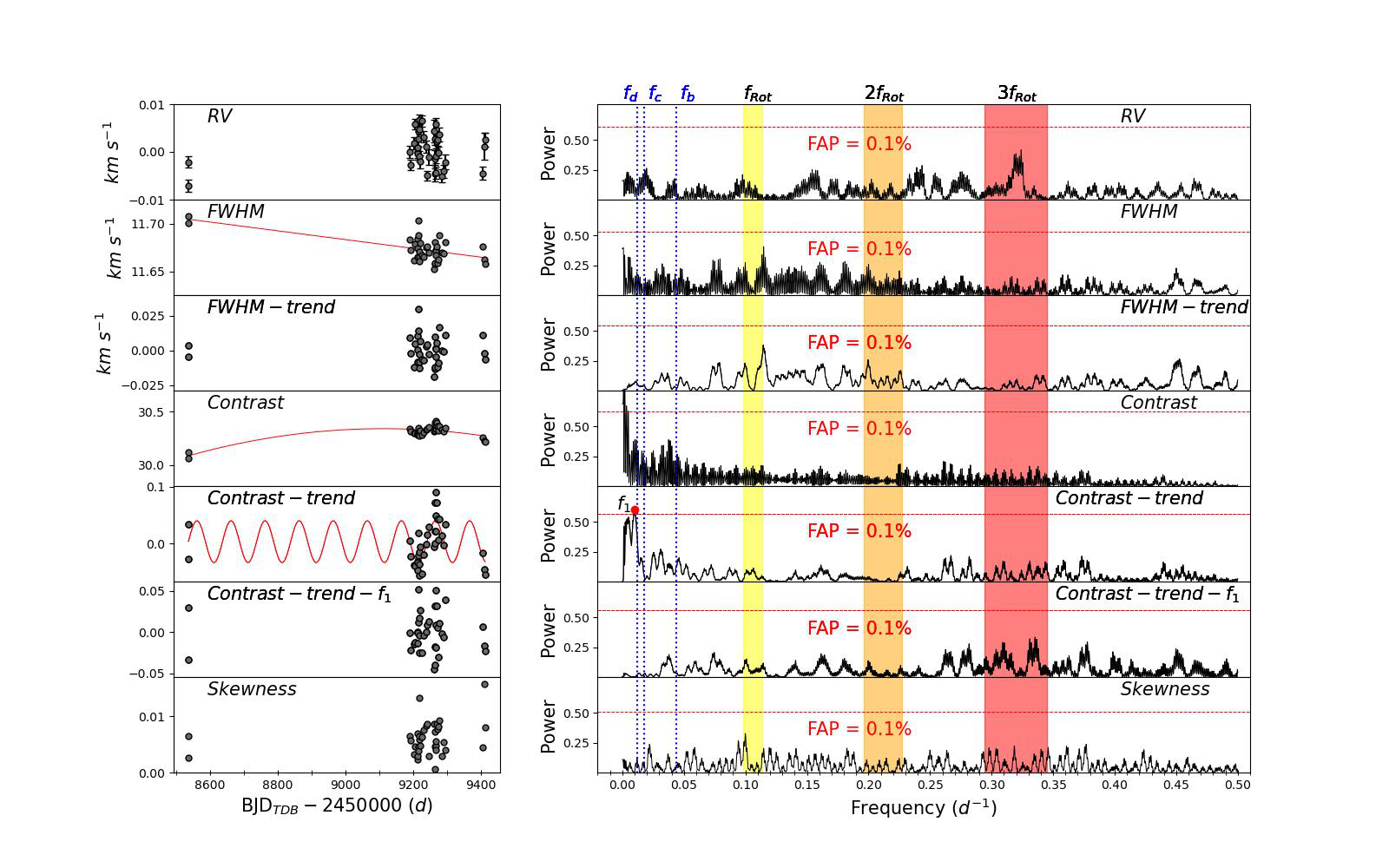}
        \caption{Time series (left column) and GLS periodograms (right column) of the RV measurements and CCF profile variation diagnostics, as extracted from the HARPS spectra of TOI-282. On the left, the red curves in the second, fourth, and fifth panels represent the linear and quadratic trends, and the sine function, as derived from the best fit to the highest peak detected in the corresponding GLS periodogram. On the right, the horizontal dotted red lines mark the 0.1\% false alarm probability (FAP). The shaded areas in yellow, orange, and red mark the rotation frequency $f_\mathrm{rot}$, and its first and second harmonics ($2f_\mathrm{rot}$ and $3f_\mathrm{rot}$), along with their 1$\sigma$ confidence intervals. The transit frequencies $f_\mathrm{b}$, $f_\mathrm{c}$, and $f_\mathrm{d}$ of the three planets are marked with vertical blue dotted lines.}
        \label{fig: Periodograms1}
\end{figure*}

\begin{figure*}[]
        \centering
        \includegraphics[width=0.8\linewidth, trim={2.6cm, 1.3cm, 3.75cm, 2.25cm}, clip]{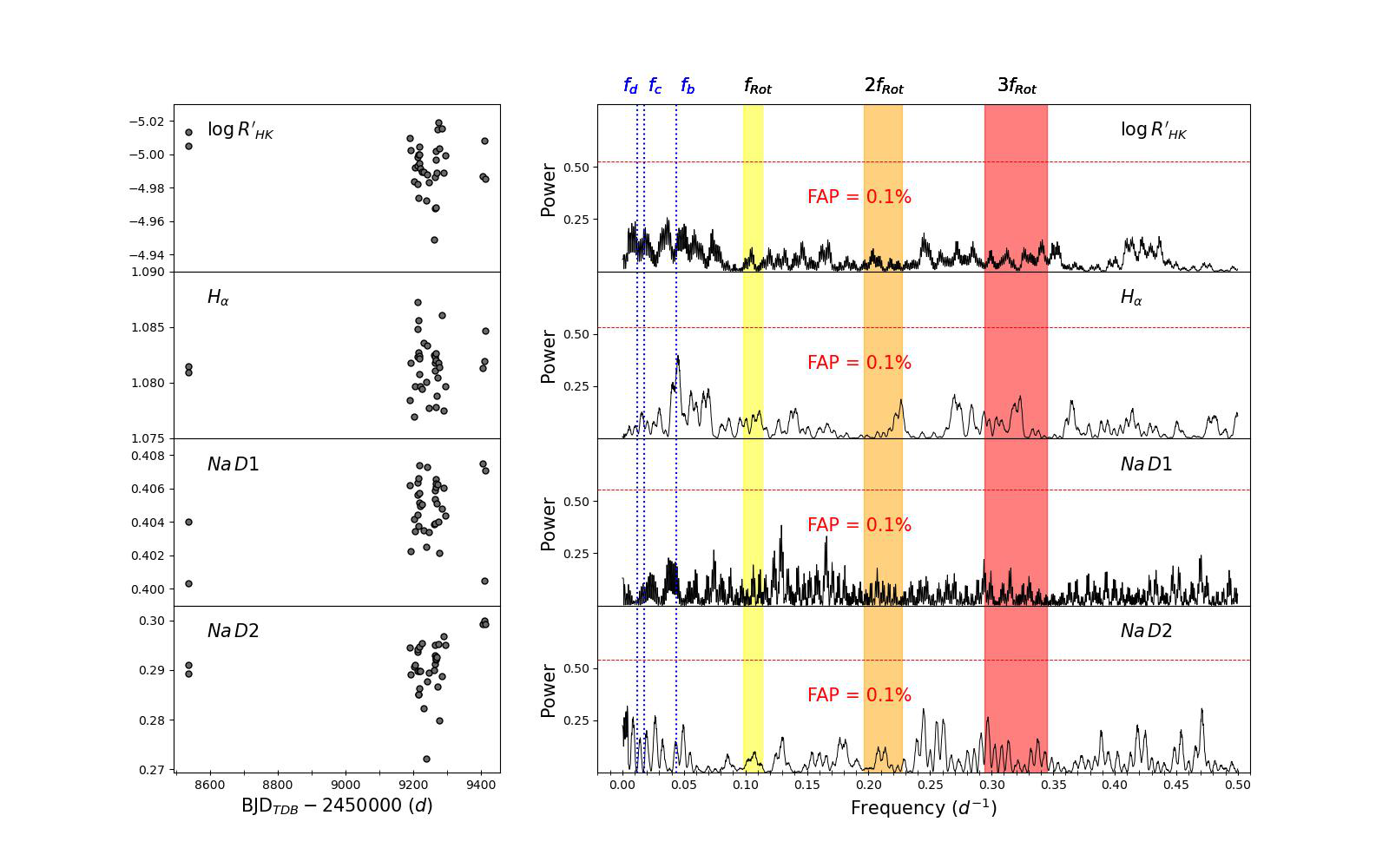}
        \caption{Time series (left column) and GLS periodograms (right column) of the HARPS activity indicators, namely \logrhk, H$_{\alpha}$, Na\,D1, and Na\,D2. On the right, the horizontal dotted red lines mark the 0.1\% false alarm probability (FAP). The shaded areas in yellow, orange, and red mark the rotation frequency $f_\mathrm{rot}$, and its first and second harmonics ($2f_\mathrm{rot}$ and $3f_\mathrm{rot}$), along with their 1$\sigma$ intervals. The transit frequencies $f_\mathrm{b}$, $f_\mathrm{c}$, and $f_\mathrm{d}$ of the three planets are marked with vertical blue dotted lines.}
        \label{fig: Periodograms2}
\end{figure*}

\begin{figure*}[]
        \centering
        \includegraphics[width=0.8\linewidth, trim={2.6cm, 1.3cm, 3.75cm, 2.25cm}, clip]{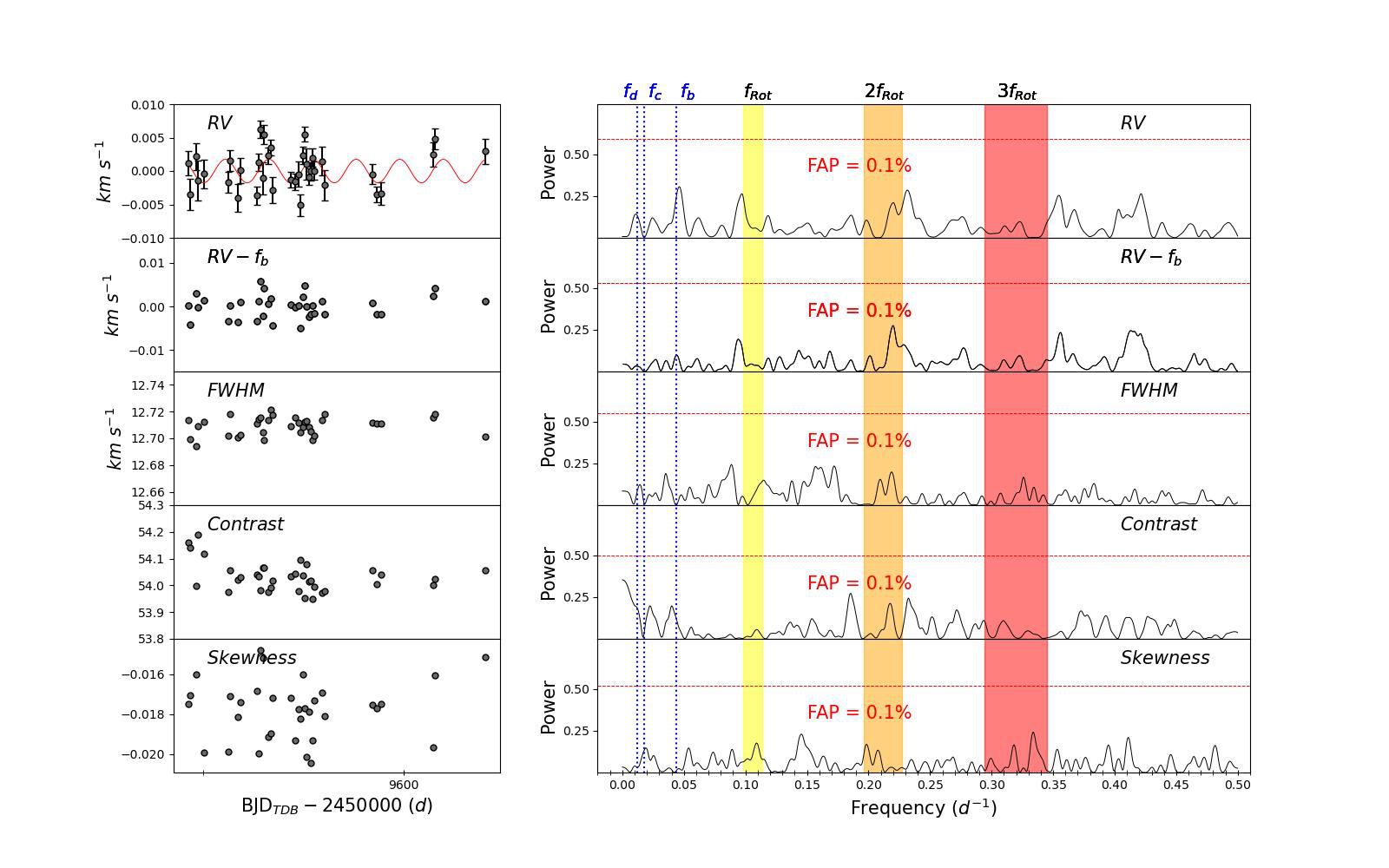}
        \caption{Time series (left column) and GLS periodograms (right column) of the RV measurements and CCF profile variation diagnostics, as extracted from the ESPRESSO 2nd data set of TOI-282. On the left, the red curve in the first panel represents the best-fitting sine function, as derived by fixing period $P_b$ and mid-time of reference transit $T_{0,b}$ of TOI-282\,b, while fitting for RV amplitude and offset. On the right, the horizontal dotted red lines mark the 0.1\% false alarm probability (FAP). The shaded areas in yellow, orange, and red mark the rotation frequency $f_\mathrm{rot}$, and its first and second harmonics ($2f_\mathrm{rot}$ and $3f_\mathrm{rot}$), along with their 1$\sigma$ intervals. The transit frequencies $f_\mathrm{b}$, $f_\mathrm{c}$, and $f_\mathrm{d}$ of the three planets are marked with vertical blue dotted lines.}
        \label{fig: Periodograms3}
\end{figure*}

We conducted a frequency analysis of the HARPS and ESPRESSO RV measurements, CCF profile variation diagnostics, and activity indicators to search for the Doppler reflex motions induced by the three transiting planets and detect potential signals resulting from stellar activity \citep[see, e.g.][]{Hatzes2010, Dumusque2011, Gandolfi2017, Serrano2022, Goffo2023}. We did not include the ESPRESSO 1st set in this part of the analysis, as it contains only 8 points (Sect.~\ref{ssec: ESPRESSO spectroscopic data}). We computed the generalised Lomb-Scargle periodograms \citep[GLS;][]{Zechmeister2009} in the frequency range  0.0\,-\,0.5~d$^{-1}$ (Figs.~\ref{fig: Periodograms1}, \ref{fig: Periodograms2}, and \ref{fig: Periodograms3}), and assessed the significance of the peaks by determining their false alarm probability (FAP), i.e. the probability that noise could generate a peak with power equal to or exceeding the one observed in the periodogram of the data. To account for possible non-Gaussian noise, we estimated the FAP using the bootstrap randomisation method \citep{Murdoch1993, Kuerster1997}. Briefly, we created $10^5$ mock periodograms by randomly shuffling the data points and respective error bars, while keeping the timestamps fixed. We defined the FAP as the fraction of those mock periodograms whose highest power is equal to or exceeds the power of the real data in the frequency interval 0.0\,-\,0.5~d$^{-1}$. We considered a peak to be significant if its FAP\,$<$\,0.1\% (horizontal red dotted lines, Figs.~\ref{fig: Periodograms1}, \ref{fig: Periodograms2}, and \ref{fig: Periodograms3}).       

The periodogram of the HARPS RVs (Fig.~\ref{fig: Periodograms1}, first row) shows no peaks with a FAP smaller than 0.1\%. Yet, an excess of power is found at the second harmonic of the stellar rotation frequency estimated in Sect.~\ref{sec: Stellar characterization}. We found that the power spectrum of the contrast (fourth row) exhibits a long-term trend, which might be attributed to long-term stellar variability, resulting in a significant excess of power at frequencies lower than the frequency resolution of our HARPS data\footnote{The frequency resolution is defined as the inverse of the time baseline. The HARPS and the ESPRESSO 2nd sets cover a baseline of about 877.3~d and 147.7~d, respectively, which yield a frequency resolution of $\sim$0.0011\,d$^{-1}$ and $\sim$0.0068\,d$^{-1}$.}. We modelled the long-period signal as a quadratic trend and subtracted the best-fitting parabola from the time series of the contrast. The periodogram of the residuals (fifth row) shows a significant signal at $\sim$0.01~d$^{-1}$ ($\sim$100~d), potentially resulting from spot evolution, which we also subtracted by fitting a sine function. Although the periodogram of the contrast residuals -- as derived from the second iteration --  does not display any significant signal, the highest peaks are found close to the second harmonic of the stellar rotation frequency (sixth row). Similarly, a possible long-period trend is also seen in the FWHM (second row), which, however, does not result in a significant peak. An attempt to remove this signal using a linear fit provides non-significant power at a frequency consistent with the stellar rotation frequency (third row). The periodograms of the HARPS activity indicators (Fig.~\ref{fig: Periodograms2}, second row) show no significant peaks. We note that, although the power spectrum of H$_{\alpha}$ displays a peak close to the transit frequency of the innermost planet (TOI-282\,b), its FAP is $\sim$9\%, making the signal not significant.

None of the periodograms of the ESPRESSO 2nd time series (Fig.~\ref{fig: Periodograms3}) displays peaks whose FAP is less than 0.1\%. Nonetheless, the periodograms of the CCF profile diagnostics exhibit peaks at frequencies consistent with the stellar rotation frequency and its harmonics. We also found that the periodogram of the ESPRESSO 2nd RVs (first row) shows a peak at the transit frequency of the inner planet TOI-282\,b. We subtracted this signal from the ESPRESSO 2nd RVs by performing a least-squares sine-fit to the amplitude and offset, while fixing the period and phase to those reported by \citetalias{Dransfield12022}. Although we found no additional signals at the transit frequency of the other two planets, the periodogram of the RV residuals shows peaks at the first harmonic of the stellar rotation frequency.

\section{Additional tables and figures}
\label{sec: Appendix}

\begin{table*}[]
\centering
\caption{Radial velocity measurements, full width at half maximum (FWHM), contrast, and skewness extracted from the cross-correlation functions of TOI-282's HARPS spectra.}
\label{table:HARPS_RV_Table}
\begin{tabular}{cccccrccrcccc}
\hline
\hline
\noalign{\smallskip}
BJD$_\mathrm{TDB}$ & RV & eRV & RV$_{\text{detrended}}$ & eRV$_{\text{detrended}}$ & FWHM  & Contrast ($A$) & Skewness ($\gamma$)& T$_\mathrm{exp}$ & S/N \\
   $-$2\,450\,000  & (\kms) & (\kms) & (\kms) & (\kms) & (\kms) & (\%)   &    &    (s)           &     \\
\noalign{\smallskip}
\hline
\noalign{\smallskip}
 8535.556886 & $-0.0071$ & 0.0013 & $-0.0005$ & 0.0016 & 11.7081 & 30.1189 & 0.0027 & 1800 & 111.7 \\
 8536.522947 & $-0.0021$ & 0.0012 & $-0.0002$ & 0.0015 & 11.7002 & 30.0595 & 0.0064 & 1800 & 120.9 \\
 9189.671091 & $-0.0001$ & 0.0013 & $-0.0020$ & 0.0016 & 11.6836 & 30.3421 & 0.0065 & 1800 & 103.8 \\
 9192.639491 & $-0.0028$ & 0.0011 & $-0.0016$ & 0.0014 & 11.6724 & 30.3142 & 0.0057 & 1800 & 120.6 \\
\ldots & \ldots & \ldots & \ldots & \ldots & \ldots & \ldots & \ldots \\ 
\noalign{\smallskip}
\hline
\end{tabular}
\tablefoot{Barycentric Julian dates are given in barycentric dynamical time \citep[BJD$_\mathrm{TDB}$;][]{Eastman2010}. The last two columns provide the exposure time and S/N per pixel at 550\,nm. The detrended dataset RV$_{\text{detrended}}$ was derived following the procedure outlined in Sect.~\ref{ssec: Radial velocity extraction and detrending}. The errors associated with the detrended RV measurements account for the jitter. The entire RV data set is available in a machine-readable table at the Strasbourg astronomical Data Center (CDS).}
\end{table*}

\begin{table*}[]
\centering
\caption{Ca\,{\sc ii} H\,\&\,K lines chromospheric index (\logrhk) and Na\,D1, Na\,D2, and H$_{\alpha}$ lines activity indicators extracted from the TOI-282's HARPS spectra.}
\label{table:HARPS_ActivityIndicators_Table}
\begin{tabular}{cccccc}
\hline
\hline
\noalign{\smallskip}
BJD$_\mathrm{TDB}$ & \logrhk & $\sigma$\,\logrhk &  Na D1  & Na D2 & H$\alpha$ \\
   $-$2\,450\,000  &         &                   &   \AA   & \AA   &  \AA      \\
\noalign{\smallskip}
\hline
\noalign{\smallskip}
 8535.556886 & $-5.0053$ & 0.0088 & 0.4003 & 0.2910 & 1.0815 \\
 8536.522947 & $-5.0133$ & 0.0081 & 0.4040 & 0.2893 & 1.0810 \\
 9189.671091 & $-5.0098$ & 0.0069 & 0.4062 & 0.2946 & 1.0784 \\
 9192.639491 & $-5.0025$ & 0.0051 & 0.4023 & 0.2891 & 1.0818 \\
\ldots & \ldots & \ldots & \ldots & \ldots & \ldots \\ 
\noalign{\smallskip}
\hline
\end{tabular}
\tablefoot{Barycentric Julian dates are given in barycentric dynamical time \citep[BJD$_\mathrm{TDB}$;][]{Eastman2010}. The entire data set is available in a machine-readable table at the Strasbourg astronomical Data Center (CDS).}
\end{table*}

\begin{table*}[]
\centering
\caption{Radial velocity measurements, full width at half maximum (FWHM), contrast, and skewness extracted from the cross-correlation functions of the TOI-282's ESPRESSO spectra.}
\label{table:ESPRESSO_RV_Table}
\begin{tabular}{cccccrccrcccc}
\hline
\hline
\noalign{\smallskip}
BJD$_\mathrm{TDB}$ & RV & eRV & RV$_{\text{detrended}}$ & eRV$_{\text{detrended}}$ & FWHM  & Contrast (A) & Skewness ($\gamma$) & T$_\mathrm{exp}$ & S/N \\
   $-$2\,450\,000  & (\kms) & (\kms) & (\kms) & (\kms) & (\kms) & (\%)   &    &    (s)           &     \\
\noalign{\smallskip}
\hline
\noalign{\smallskip}
\multicolumn{3}{l}{ESPRESSO 1st set} \\
\noalign{\smallskip}
 8771.715834 & $-0.0061$ & 0.0022 & $-0.0019$ & 0.0026 & 12.7080 & 54.1188 & $-0.0206$ & 1115 &  88.2 \\
 8776.730177 &  0.0011 & 0.0022 & $0.0044$ & 0.0026 & 12.6990 & 53.9773 & $-0.0198$ & 1115 &  88.5 \\
 8788.726805 & $-0.0010$ & 0.0023 & $0.0007$ & 0.0027 &  12.7019 & 54.0500 & $-0.0182$ & 1115 &  82.5 \\
 8804.620886 & $-0.0079$ & 0.0017 & $-0.0046$ & 0.0022 & 12.7019 & 53.9437 & $-0.0198$ & 1115 & 114.7 \\
\ldots & \ldots & \ldots & \ldots & \ldots & \ldots & \ldots & \ldots \\ 
\noalign{\smallskip}
\hline
\noalign{\smallskip}
\multicolumn{3}{l}{ESPRESSO 2nd set} \\
\noalign{\smallskip}
 9492.867880 &  0.0011 & 0.0018 & $0.0026$ & 0.0019 & 12.7134 & 54.1612 & $-0.0175$ &  935 & 105.5 \\
 9493.733784 & $-0.0035$ & 0.0024 & $-0.0002$ & 0.0024 & 12.6995 & 54.1398 & $-0.0170$ &  935 &  79.3 \\
 9496.727138 &  0.0022 & 0.0020 & $0.0003$ & 0.0021 & 12.6939 & 53.9965 & -0.0160 &  935 &  95.1 \\
 9497.753690 & $-0.0014$ & 0.0030 & $-0.0011$ & 0.0030 & 12.7090 & 54.1882 & $-0.0220$ &  935 &  63.1 \\
\ldots & \ldots & \ldots & \ldots & \ldots & \ldots & \ldots & \ldots \\ 
\noalign{\smallskip}
\hline
\end{tabular}
\tablefoot{Barycentric Julian dates are given in barycentric dynamical time \citep[BJD$_\mathrm{TDB}$;][]{Eastman2010}. We split the data into ESPRESSO 1st and ESPRESSO 2nd sets, and treat the two sets independently to account for a potential RV offset that may have resulted from the instrument refurbishment performed between 2019 and 2020. The last two columns provide the exposure time and S/N per pixel at 550\,nm. The detrended dataset RV$_{\text{detrended}}$ was derived following the procedure outlined in Sect.~\ref{ssec: Radial velocity extraction and detrending}. The errors associated with the detrended measurements account for the jitter. The entire RV data set is available in a machine-readable table at the Strasbourg astronomical Data Center (CDS)}
\end{table*}

\begin{table}[]
    \small\centering\renewcommand{\arraystretch}{1.2}
    \caption{Transit times of TOI-282\,b from the LCs + RVs \texttt{MCMCI} analysis with TTVs}
    \begin{tabular}{c c c r}
    \hline\hline
    $T_{b,\,j}$ ($\mathrm{BJD_{TDB}}$) & $\sigma_{T0}$ (days) & O$-$C (minutes) & Telescope \\
    \hline
$2\,458\,344.8135$ & $0.024$ & $-2.17$ & TESS \\
$2\,458\,390.5660$ & $0.023$ & $-44.79$ & TESS \\
$2\,458\,413.4903$ & $0.016$ & $3.09$ & TESS \\
$2\,458\,482.1727$ & $0.017$ & $16.44$ & TESS \\
$2\,458\,505.0461$ & $0.011$ & $-9.06$ & TESS \\
$2\,458\,527.9499$ & $0.024$ & $9.31$ & TESS \\
$2\,458\,550.8404$ & $0.013$ & $8.54$ & TESS \\
$2\,458\,573.7825$ & $0.024$ & $81.97$ & TESS \\
$2\,458\,619.4928$ & $0.019$ & $-21.43$ & TESS \\
$2\,458\,665.2832$ & $0.013$ & $-9.45$ & TESS \\
$2\,459\,054.4329$ & $0.026$ & $-6.94$ & TESS \\
$2\,459\,077.3150$ & $0.029$ & $-19.78$ & TESS \\
$2\,459\,123.0878$ & $0.016$ & $-33.13$ & TESS \\
$2\,459\,146.0117$ & $0.023$ & $14.10$ & TESS \\
$2\,459\,168.9013$ & $0.012$ & $11.99$ & TESS \\
$2\,459\,237.5973$ & $0.024$ & $44.89$ & TESS \\
$2\,459\,329.1261$ & $0.014$ & $-6.08$ & TESS \\
$2\,459\,352.0039$ & $0.037$ & $-25.12$ & TESS \\
$2\,459\,970.0749$ & $0.016$ & $-7.16$ & TESS \\
$2\,459\,992.9855$ & $0.018$ & $21.04$ & TESS \\
$2\,460\,038.7394$ & $0.031$ & $-19.53$ & TESS \\
$2\,460\,130.3380$ & $0.014$ & $29.90$ & TESS \\
$2\,460\,176.0815$ & $0.017$ & $-25.68$ & TESS \\
$2\,460\,702.5884$ & $0.012$ & $-7.43$ & TESS \\
    \hline
    \end{tabular}
    \tablefoot{The O$-$C values (third column) are computed with respect to the linear ephemeris: \\ $T_\mathrm{ref} = 2\,458\,344.8150 \pm0.0053\, \mathrm{BJD_{TDB}}$\\ $P_\mathrm{lin} = 22.89105\pm 0.00011 $~days. \\ The transit times are given in the $\mathrm{BJD_{TDB}}$ standard \citep{Eastman2010}; the second column reports the associated 1-$\sigma$ error.}
    \label{table:T0_b}
\end{table}

\begin{table}[]
    \small\centering\renewcommand{\arraystretch}{1.2}
    \caption{Transit times of TOI-282\,c from the LCs + RVs \texttt{MCMCI} analysis with TTVs}
    \begin{tabular}{c c c r}
    \hline\hline
    $T_{c,\,j}$ ($\mathrm{BJD_{TDB}}$) & $\sigma_{T0}$ (days) & O$-$C (minutes) & Telescope \\
    \hline
$2\,458\,337.2107$ & $0.0062$ & $-104.47$ & TESS \\
$2\,458\,393.2147$ & $0.0038$ & $-95.56$ & TESS \\
$2\,458\,449.2233$ & $0.0029$ & $-80.14$ & TESS \\
$2\,458\,505.2422$ & $0.0042$ & $-49.94$ & TESS \\
$2\,458\,561.2524$ & $0.0047$ & $-32.10$ & TESS \\
$2\,458\,617.2629$ & $0.0046$ & $-13.94$ & TESS \\
$2\,458\,673.2726$ & $0.0079$ & $3.07$ & TESS \\
$2\,459\,065.3157$ & $0.0036$ & $86.53$ & TESS \\
$2\,459\,121.3099$ & $0.0038$ & $81.27$ & TESS \\
$2\,459\,233.3195$ & $0.0051$ & $101.22$ & TESS \\
$2\,459\,289.3181$ & $0.0067$ & $102.35$ & TESS \\
$2\,459\,345.3289$ & $0.005$ & $120.86$ & TESS \\
$2\,460\,017.2774$ & $0.0091$ & $83.49$ & TESS \\
$2\,460\,073.2459$ & $0.0068$ & $41.20$ & TESS \\
$2\,460\,129.2423$ & $0.004$ & $39.08$ & TESS \\
$2\,460\,185.2284$ & $0.0062$ & $22.14$ & TESS \\
$2\,460\,689.0951$ & $0.0026$ & $-142.25$ & TESS \\
$2\,460\,745.0787$ & $0.0064$ & $-162.80$ & TESS \\
    \hline
    \end{tabular}
    \tablefoot{The O$-$C values (third column) are computed with respect to the linear ephemeris: \\ $T_\mathrm{ref} = 2\,458\,337.283 \pm0.024\, \mathrm{BJD_{TDB}}$\\ $P_\mathrm{lin} = 55.9978\pm 0.0011 $~days. \\ The transit times are given in the $\mathrm{BJD_{TDB}}$ standard \citep{Eastman2010}; the second column reports the associated 1-$\sigma$ error.}
    \label{table:T0_c}
\end{table}

\begin{table*}[]
    \small\centering\renewcommand{\arraystretch}{1.2}
    \caption{Transit times of TOI-282\,d from the LCs + RVs \texttt{MCMCI} analysis with TTVs}
    \begin{tabular}{c c c r}
    \hline\hline
    $T_{d,\,j}$ ($\mathrm{BJD_{TDB}}$) & $\sigma_{T0}$ (days) & O$-$C (minutes) & Telescope \\
    \hline
$2\,458\,355.6757$ & $0.014$ & $154.37$ & TESS \\
$2\,458\,439.9146$ & $0.012$ & $81.06$ & TESS \\
$2\,458\,524.1570$ & $0.01$ & $12.67$ & TESS \\
$2\,458\,608.4130$ & $0.013$ & $-36.09$ & TESS \\
$2\,459\,282.5086$ & $0.022$ & $-357.54$ & TESS \\
$2\,459\,366.7951$ & $0.016$ & $-362.37$ & TESS \\
$2\,460\,715.8058$ & $0.014$ & $174.56$ & TESS \\
    \hline
    \end{tabular}
    \tablefoot{The O$-$C values (third column) are computed with respect to the linear ephemeris: \\ $T_\mathrm{ref} = 2\,458\,355.5684 \pm0.0062\, \mathrm{BJD_{TDB}}$\\ $P_\mathrm{lin} = 84.28985\pm 0.00055 $~days. \\ The transit times are given in the $\mathrm{BJD_{TDB}}$ standard \citep{Eastman2010}; the second column reports the associated 1-$\sigma$ error.}
    \label{table:T0_d}
\end{table*}

\begin{table*}[]
    \centering\small\centering\renewcommand{\arraystretch}{1.3}
    \caption{Posteriors and derived orbital parameters (MAP and HDI) for TOI-282\,b, c, and d obtained from the dynamical analysis with \trades.}
    \begin{tabular}{l c c r}
    \hline\hline
    Parameter & Unit & Prior & MAP (HDI$\pm 1\sigma$) \\
    \hline
    \emph{TOI-282\,b} \rule{0pt}{12pt} & & & \\
    &&& \\
    Orbital Period ($P$) & [days] & \unif{21.5}{23.5} & $22.89076_{-0.00015}^{+0.00017}$  \\
    Mass ($m_{\mathrm{p}}$) & [$M_{\oplus}$] & -- &  $6.7_{-0.8}^{+1.7}$ \\
    Eccentricity ($e$) & -- & --  & $0.0071_{-0.0071}^{+0.0475}$ \\
    Argument of Periastron ($\omega$) & [deg] & -- & $-93_{-54}^{+54}$ \\
    Mean Longitude ($\lambda$) & [deg] & \unif{0}{360} & $147_{-5}^{+1}$ \\
    Mean Anomaly ($M_{\mathrm{A}}$) & [deg] & -- & $60_{-90}^{+90}$ \\

    \hline
    
    \emph{TOI-282\,c} \rule{0pt}{12pt} & & & \\
    &&& \\
    Orbital Period ($P$) & [days] & \unif{54.5}{57.5} & $56.00633_{-0.00083}^{+0.00166}$ \\
    Mass ($m_{\mathrm{p}}$) & [$M_{\oplus}$] & -- & $10.0_{-2.0}^{+1.0}$ \\
    Eccentricity ($e$) & -- & -- & $0.0039_{-0.0037}^{+0.0041}$  \\
    Argument of Periastron ($\omega$) & [deg] & -- & $39_{-6}^{+73}$ \\
    Mean Longitude ($\lambda$) & [deg] & \unif{0}{360} & $268.31_{-0.15}^{+0.51}$ \\
    Mean Anomaly ($M_{\mathrm{A}}$) & [deg] & -- &  $50_{-72}^{+7}$ \\

    \hline
    
    \emph{TOI-282\,d} \rule{0pt}{12pt} & & & \\
    &&& \\
    Orbital Period ($P$) & [days] &  \unif{83}{87} & $84.2721_{-0.0053}^{+0.0039}$ \\
    Mass ($m_{\mathrm{p}}$) & [$M_{\oplus}$] & -- & $5.8_{-1.1}^{+0.9}$ \\
    Eccentricity ($e$) & -- & -- & $0.0054_{-0.0036}^{+0.0027}$ \\
    Argument of Periastron ($\omega$) & [deg] & -- & $-72_{-27}^{+42}$ \\
    Mean Longitude ($\lambda$) & [deg] & \unif{0}{360} & $190.04_{-0.30}^{+0.23}$ \\
    Mean Anomaly ($M_{\mathrm{A}}$) & [deg] & -- &  $82_{-43}^{+27}$ \\

    \hline
    \end{tabular}
    \label{table:fit_TRADES_orbital_parameters}
    \tablefoot{The symbols $\mathcal{U}$, $\mathcal{G}$, and $\mathcal{N}^{+}$ refer to uniform, Gaussian, and half-Gaussian distributions, respectively.}

\end{table*}

\begin{table*}[]

    \caption{Polynomial detrending baseline models.}
    \makebox[\textwidth][c]{%
    \parbox[t]{0.35\textwidth}{
    \centering\small\renewcommand{\arraystretch}{1.3}
    \begin{tabular}{l c c}
    \hline\hline
    \makecell{$T_0$ [BJD$_{\text{TBD}}-2\,450\,000$]} & Planet & Detrending model \\
    \hline
    8337.2106 & c & -- \\
    8344.8129 & b & -- \\
    8355.6744 & d & $\Delta y^2$ \\
    8390.5663 & b & t$^2$ \\
    8393.2147 & c & -- \\
    8413.4904 & b & -- \\
    8436.3172 & b & -- \\
    8439.9165 & d & -- \\
    8449.2233 & c & -- \\
    8459.2706 & b & $\Delta x^2$ \\
    8482.1724 & b & $\Delta y^1$ \\
    8505.0478 & b & t$^3$ \\
    8505.2423 & c & -- \\
    8524.1579 & d & -- \\
    8527.9507 & b & -- \\
    8550.8334 & b & -- \\
    8561.2526 & c & -- \\
    8573.7809 & b & -- \\   
    8608.4129 & d & -- \\
    8617.2631 & c & -- \\
    8619.4950 & b & -- \\
    8665.2876 & b & -- \\
    8673.2721 & c & -- \\
    9054.4329 & b & -- \\
    9065.3155 & c & -- \\
    9077.3087 & b & -- \\
    9121.3102 & c & -- \\
    9123.0901 & b & -- \\
    9146.0351 & b & -- \\
    9168.9019 & b & $\Delta y^1$ \\
    9233.3195 & c & -- \\
    9237.6020 & b & -- \\
    \hline
    \end{tabular}%
    }%

            
    \parbox[t]{0.62\textwidth}{%
    \centering\small\renewcommand{\arraystretch}{1.3}
    \begin{tabular}{l c c}
    \hline\hline
    \makecell{$T_0$ [BJD$_{\text{TBD}}-2\,450\,000$]} & Planet & Detrending model \\
    \hline
    9260.4520 & b & -- \\
    9282.5072 & d & $\Delta y^1$ \\
    9283.3496 & b & -- \\
    9289.3180 & c & -- \\
    9329.1223 & b & -- \\
    9345.3287 & c & -- \\
    9352.0121 & b & $\Delta y^1$ \\
    9366.8022 & d & t$^1$ \\
    9375.0266 & b & --\\
    9970.0776 & b & -- \\
    9992.9890 & b & -- \\
    10015.7343 & b & -- \\
    10017.2770 & c & -- \\
    10038.7653 & b & -- \\
    10041.5917 & d & -- \\
    10061.8997 & b & -- \\
    10073.2459 & c & -- \\
    10107.4239 & b & -- \\
    10129.2429 & c & -- \\
    10130.3397 & b & -- \\
    10176.0838 & b & -- \\ 
    10185.2292 & c & -- \\
    10198.9941 & b & -- \\
    10689.0985 & c & $\text{t}^4 + \Delta x^1 + \Delta y^1 + (x-y)^3$ \\
    10702.5869 & b & t$^1$ \\
    10715.8308 & d & t$^1$ \\
    10745.0779 & c & t$^4$ \\

    \hline
    \hline

    Time series & \multicolumn{2}{c}{Detrending model} \\
    \hline
    HARPS & \multicolumn{2}{c}{$t^2+ \text{FWHM}^2 + \gamma^1 + A^3 + H_{\alpha}^1$} \\
    ESPRESSO 1st set & \multicolumn{2}{c}{$\gamma^1$} \\
    ESPRESSO 2nd set & \multicolumn{2}{c}{$t^2 + \text{FWHM}^1 + \gamma^2 + A^1$} \\
    \hline
    \end{tabular}%
    } 
    }
    \label{table: detrending polynoms}
\end{table*}

\begin{figure*}
    \centering
    \includegraphics[width=0.33\textwidth]{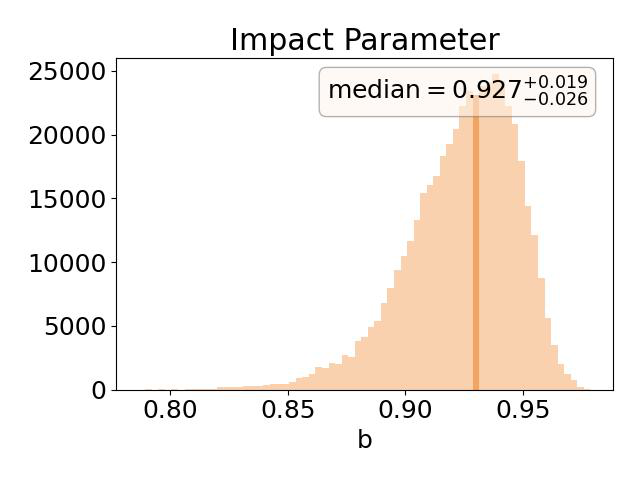}
    \includegraphics[width=0.33\textwidth]{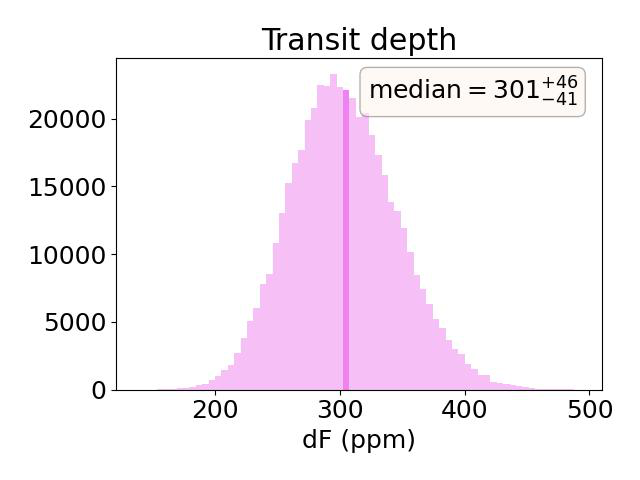} 
    \includegraphics[width=0.33\textwidth]{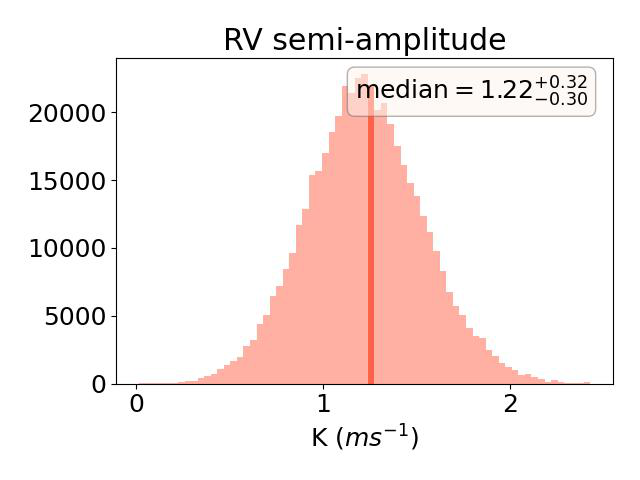}
    \caption{Posterior distributions of the relevant model parameters outputted by the \texttt{MCMCI} for TOI-282\,b: impact parameter (left panel), transit depth (middle panel), and RV semi-amplitude (right panel).}
    \label{fig: posteriorlinb}
\end{figure*}

\begin{figure*}
    \centering
    \includegraphics[width=0.33\textwidth]{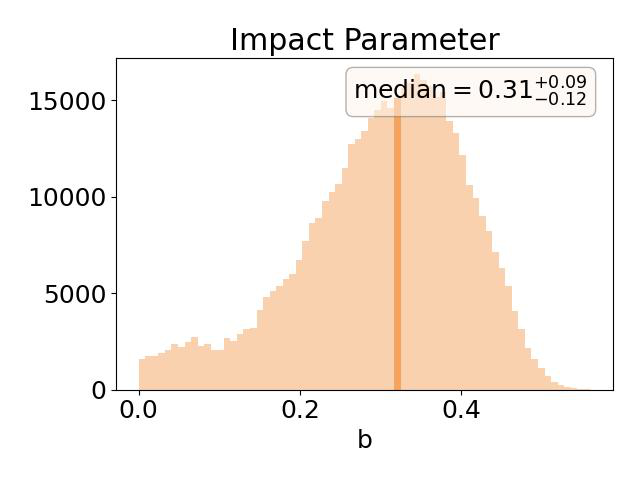}
    \includegraphics[width=0.33\textwidth]{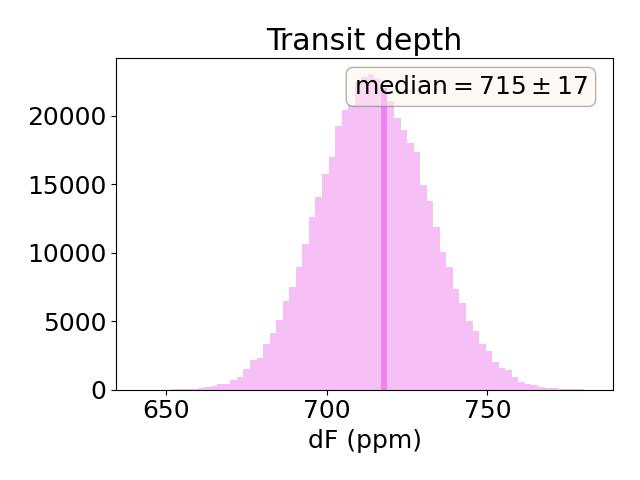} 
    \includegraphics[width=0.33\textwidth]{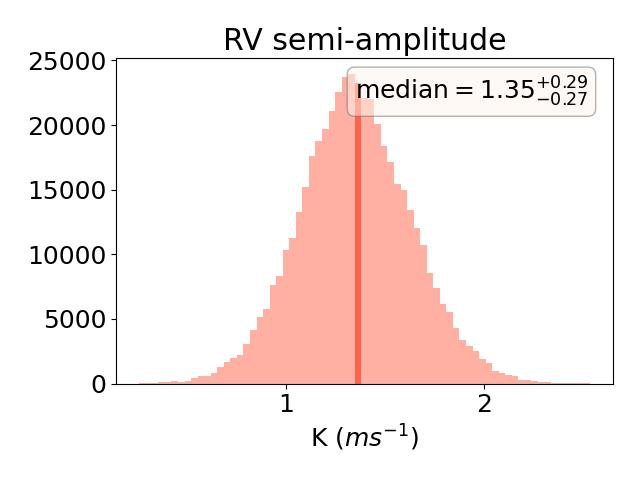}
    \caption{Posterior distributions of the relevant parameters outputted by the \texttt{MCMCI} for TOI-282\,c: impact parameter (left panel), transit depth (middle panel), and RV semi-amplitude (right panel).}
    \label{fig: posteriorlinc}
\end{figure*}

\begin{figure*}
    \centering
    \includegraphics[width=0.33\textwidth]{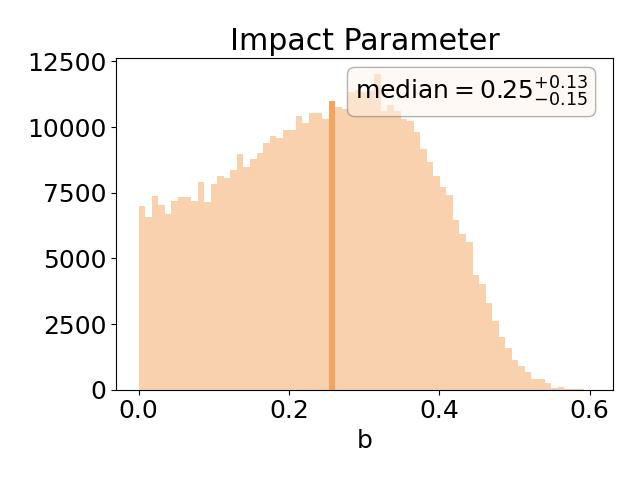}
    \includegraphics[width=0.33\textwidth]{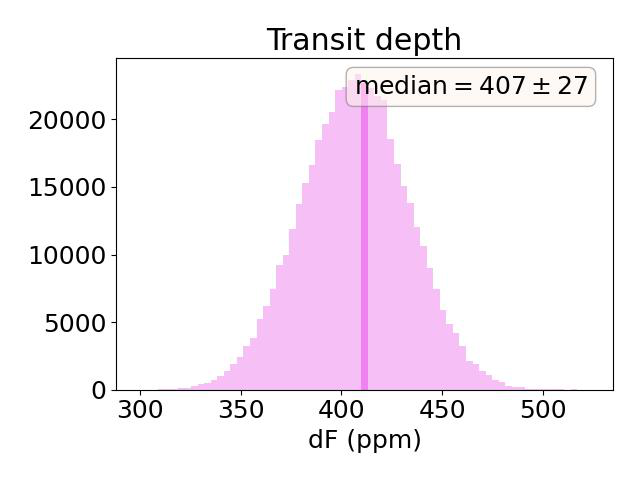} 
    \includegraphics[width=0.33\textwidth]{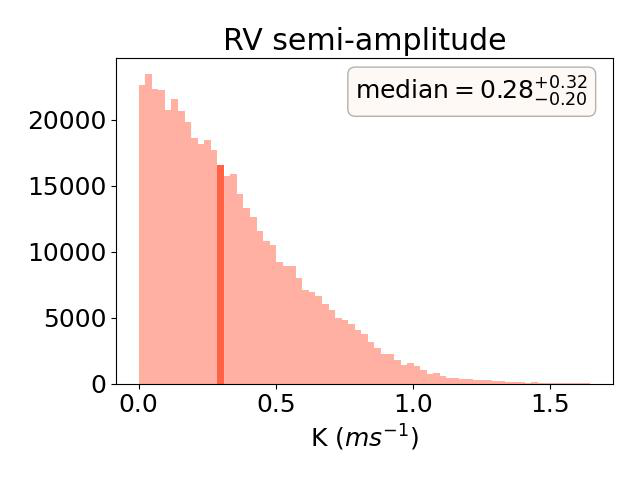}
    \caption{Posterior distributions of the relevant parameters outputted by the \texttt{MCMCI} for TOI-282\,d: impact parameter (left panel), transit depth (middle panel), and RV semi-amplitude (right panel).}
    \label{fig: posteriorlind}
\end{figure*}

\end{appendix}

\end{document}